\documentclass[preprint]{aastex}
\usepackage{psfig}

\newcommand{\kms}{km~s$^{-1}$}
\newcommand{\vrot}{V$_{\rm rot}$}

\begin{document}

\slugcomment{}

\title{Rotationally Supported Virgo Cluster Dwarf Elliptical Galaxies: \\ Stripped Dwarf Irregular Galaxies?}

\author{Liese van~Zee}
\affil{Astronomy Department, Indiana University, 727 E 3rd St, Bloomington, IN 47405}
\email{vanzee@astro.indiana.edu}
\author{Evan D. Skillman}
\affil{Astronomy Department, University of Minnesota, 116 Church St. SE, Minneapolis, MN 55455}
\email{skillman@astro.umn.edu}
\and
\author{Martha P. Haynes}
\affil{Center for Radiophysics and Space Research and
National Astronomy and Ionosphere Center,\altaffilmark{1} Cornell University, Ithaca, NY 14853}
\email{haynes@astro.cornell.edu}

\altaffiltext{1}{The National Astronomy and Ionosphere Center is
operated by Cornell University under a cooperative agreement with the
National Science Foundation.}

\begin{abstract}

New observations of 16 dwarf elliptical galaxies in the Virgo Cluster indicate that
at least seven dEs have significant velocity gradients along their optical major
axis, with typical rotation amplitudes of 20-30 \kms.
Of the remaining nine galaxies in this sample, 6 have velocity gradients less 
than 20 \kms kpc$^{-1}$ while the other 3 observations had too low of a 
signal--to--noise ratio to determine an accurate velocity gradient.
Typical velocity dispersions for these  galaxies are $\sim$44 $\pm$ 5 \kms, 
indicating that rotation can be a significant component of the stellar dynamics
of Virgo dEs. When corrected for the limited spatial extent of the spectral
data, the rotation amplitudes
of the rotating dEs are comparable to those of similar brightness dIs.  
Evidence for a relationship between the rotation amplitude and galaxy luminosity
is found, and, in fact, agrees well with the Tully-Fisher relation.  The 
similarity in the scaling relations of dIs and dEs implies that it is unlikely 
that dEs evolve from significantly more luminous galaxies.
These observations reaffirm the possibility that some cluster dwarf elliptical galaxies 
may be formed when the neutral gaseous medium is stripped from dwarf irregular 
galaxies in the cluster environment.
We hypothesize that several different mechanisms are involved in the creation of
the overall population of dE galaxies, and that stripping of infalling dIs
may be the dominant process in the creation of dEs in clusters like Virgo.
\end{abstract}


\section{Introduction}
Dwarf elliptical\footnote{The term ``dwarf elliptical'' is used in this paper to indicate
the most common type of low mass, gas--poor dwarf galaxy; some authors prefer to use 
``spheroidal'' and ``dwarf spheroidal''  to describe this class.} galaxies (dE) are the
most ubiquitous type of galaxy in the local universe \citep{BST88,FB94,GW94}.  Despite
their plenitude, we still do not understand fully the formation mechanism of
these low mass, gas--poor systems.
One clue to the evolution of dwarf elliptical galaxies
 is the morphology--density relation: similar to giant elliptical galaxies, dwarf 
elliptical galaxies are found primarily in high density regions such as galaxy clusters, 
compact groups, and loose groups \citep{BST88}.  However, unlike their giant cousins, the 
stellar distribution in dwarf elliptical galaxies is exponential in nature \citep{LF83,K85}, 
indicating that dwarf elliptical galaxies are not just lower mass versions of giant 
elliptical galaxies.  Thus, the evolutionary scenarios proposed for giant elliptical 
galaxies may not be relevant for the dwarf elliptical class.

While some aspects of galaxy evolution undoubtedly arise from initial conditions,
on-going evolution is likely accelerated and affected by different mechanisms
in dense cluster environments. As the nearest such location, the Virgo cluster serves
as the best laboratory for exploring the details of evolution under intracluster
conditions, providing access to spatial scales and mass and luminosity ranges that
are not achievable in more distant systems. As discussed in \citet{BST88}, the
galaxy population in Virgo is dominated by dwarf galaxies which, because of its
proximity, can be identified in large numbers. Thus, Virgo provides an ideal
setting for the study of dwarf galaxy evolution in the cluster environment.

A multitude of evolutionary scenarios have
been proposed to explain the relationship between gas--poor dwarf elliptical
galaxies and gas--rich dwarf irregular galaxies \citep[e.g.,][]{DS86,SWS87,DP88}.
The key component of these models is the gas removal process:  gas is removed either
through a ```blow--out'' following a starburst episode, or via ram pressure stripping
as the galaxy encounters the hot intergalactic medium (IGM) or intracluster medium (ICM).
 Note that the latter method provides
a natural explanation to the density-morphology relationship observed in
the dwarf galaxies of the Local Group \citep{vdB94} and in clusters \citep{BTS90},
while the former method has difficulty explaining this phenomenon.

Interestingly, if either of the above scenarios are correct, the stellar remnant (dE) 
should have the same kinematic properties as the progenitor (dI).  However, the 
ground--breaking kinematic work by \citet{BN90} on dE's in Virgo and \citet{BPN91} 
on three dE companions of M31 suggested that 
this may not be the case for dEs:  of the 5 dwarf elliptical galaxies in these two
early studies, none are supported by rotation (\vrot $<$ 20 \kms); in contrast, most 
dwarf irregular galaxies of comparable mass (or luminosity) are rotationally supported. 
Recently, however,  \citet{Me01a,Me01b}
have shown that ``tidal stirring'' -- or repeated tidal shocks suffered
by dwarf satellite galaxies -- can convert a dI into a dE by both
removing its gas and changing it into a dynamically hot galaxy.
Thus, lack of rotation in dE galaxies does not exclude the conversion of
dI to dE, but may indicate a more catastrophic evolutionary
pathway than blow--out or ram pressure stripping alone. 

The number of dwarf elliptical galaxies with spatially resolved rotation curves
is still quite small, however.  Stellar kinematics of large samples of
dwarf elliptical galaxies are needed to investigate fully kinematic
constraints on evolutionary scenarios.
Several projects with modest sample sizes have recently appeared in the 
literature \citep[e.g.,][]{PGCSBG02,GGvM02,GGvM03}.  
Here, we present the first results from a 
program to measure the stellar kinematics of dwarf galaxies;
rotation curves and velocity dispersions of sixteen dwarf elliptical 
galaxies in the Virgo Cluster are discussed in this paper.

The paper is organized as follows. The spatially resolved kinematic observations are
described in Section 2.  In Section 3, the observed kinematic properties are compared
to those of giant elliptical galaxies and dwarf irregular galaxies.
In contrast to previous studies of dEs, seven of the sixteen
galaxies in this sample have clear evidence of rotational support.    
These results are discussed in the context of dwarf galaxy evolution in Section 4.  
The conclusions of the paper are summarized in Section 5.

\section{Kinematic Observations}

\subsection{Sample Selection}
We selected a moderate sized sample of nearby dwarf elliptical galaxies for 
spatially resolved kinematic observations from the Virgo Cluster Catalog \citep[VCC,][]{VCC}.  
The target galaxies were selected based on their apparent magnitude (m$_{\rm B} < 15.5$), 
morphological classification (dE), and on their apparent ellipticity ($\epsilon > 0.25$).
The latter criterion was imposed to mitigate the affects of disk projection in the
measurement of rotational velocities, and it should be noted that its adoption
biases the sample toward objects with higher apparent rotation velocity. The implication of this
bias in terms of the galaxy evolution scenarios described in Section 4 is uncertain.
The sample was further restricted to the luminous dwarf elliptical galaxies
in the VCC with measured recessional velocities that place them within
the Virgo Cluster \citep[e.g.,][]{CGW01}, at a distance of approximately 16.1 Mpc \citep[e.g.,][]{Ke00}.
Dynamically young, Virgo is known to contain several substructural units \citep{BST87}
and appears to be elongated along the line-of-sight \citep{YFO97,SSSGH02}.
Recent observations of surface brightness fluctuations \citep{JBB04} have shown that the
dE population is stretched in depth similarly to the bright elliptical population
\citep{NT00,WB00} but not as extended as claimed previously by \citet{YC95}.
Figure \ref{fig:virgo} shows the location of the selected galaxies within the
Virgo Cluster;  the targeted galaxies were selected to include a range of environments 
within the cluster, sampling both the cluster core and the outer field.

The observational properties of the selected galaxies are tabulated in Table 
\ref{tab:global}.  Included in Table \ref{tab:global} are the morphological class 
from the VCC \citep{VCC} and 
group assignments from \citet{BPT93}.
The tabulated position angle, 
ellipticity, apparent magnitude, central surface brightness, and scale length
are derived from ELLIPSE fitting of UBVRI images obtained with the VATT 1.8m
telescope \citep{vZB04}.  We note that the 
newly derived ellipticities are in excellent agreement with those values previously 
published for VCC 543, VCC 990, VCC 1122, VCC 1308, VCC 1514, VCC 1743, and VCC 2019 
\citep{RTPL99}.  
  
\subsection{Palomar Observations}

High resolution optical spectra of the target galaxies were obtained with the 
Double Spectrograph on the 5m Palomar\footnote{Observations at the Palomar Observatory 
were made as part of a continuing cooperative agreement between Cornell University and
the California Institute of Technology.} telescope on the nights of 2001 March 19-21 
and 2002 Apr 8-10.  The 6800 \AA~dichroic was used to split the light to the two 
sides (blue and red) of the spectrograph.  Both sides of the spectrograph were equipped 
with 1200 l/mm gratings (blazed at 5000 \AA~on the blue side and blazed at 9400 \AA~on 
the red side).  The thinned TEK 1024 $\times$ 1024 CCD on the blue side of the 
spectrograph had a read noise of 8.6 e$^-$, a system gain of 2.13, and an angular scale 
of 0.624 arcsec/pix.  The thinned TEK 1024 $\times$ 1024 CCD on the red side of the 
spectrograph had a read noise of 7.5 e$^-$, a system gain of 2.0, and an angular scale 
of 0.482 arcsec/pix.  The blue side was centered at the Mg Ib triplet, with wavelength
coverage of 4800 -- 5700 \AA~and an effective resolution of 2.3 \AA~(0.88 \AA/pix).
The red side was centered at the Ca triplet, with wavelength coverage of 
8250 -- 8900 \AA~and an effective resolution of 1.6 \AA~(0.64 \AA/pix).

A summary of the observations is tabulated in Table \ref{tab:obs}.  
The 2\arcmin~long slit was oriented along the optical major axis of each galaxy. 
Since the galaxies in this sample are faint and low surface brightness, the telescope 
was centered on a nearby bright star and then offset to the galaxy position. 
The slit viewing camera provided confirmation that the slit was aligned and
centered appropriately.  To improve the sensitivity, the slit was set to a 
width of 2\arcsec~for all observations; the seeing was approximately 1\arcsec~during 
all 6 nights.  Due to the low surface brightness nature of the dE galaxies, long 
integration times (on the order of 2 -- 3 hours per pointing) 
were necessary to have sufficient signal--to--noise in the absorption lines to trace 
the stellar kinematics to large radii.

Observations of standard stars were obtained at the beginning and end of the
night.  Flux calibration was derived from observations of Feige 34, BD +75 325, 
and BD +33 2642 \citep{oke90}.  The nights were non-photometric, but generally
transparent.  At least 6 radial velocity standards (F7 - K2 giants) were 
also observed each night in order to create stellar templates for cross-correlation 
analysis.  The radial velocity standards are tabulated in Table \ref{tab:stars}. 
During the 2002 April observing run, several stars with low metallicities were targeted
in order to verify that spectral differences between the template stars
(typically solar metallicity) and the low metallicity dwarf elliptical galaxies 
did not introduce systematic errors.  As expected, we did not find
that the metallicity of the template star had a significant effect on the derived
redshift or velocity dispersion for these modest dispersion observations.

\subsection{Data Reduction}

The spectra were reduced and analyzed with the IRAF\footnote{IRAF 
is distributed by the National Optical Astronomy Observatories.} package.
The spectral reduction included bias subtraction, scattered light
corrections, and flat fielding with both twilight and dome flats.
The 2--dimensional (2-d) images were rectified based on arc lamp observations 
(blue) or night sky lines (red) and the trace of stars at different positions 
along the slit.  Confirmation of the wavelength solution was provided by 
comparing the observed heliocentric velocities of the radial velocity standards 
with those found in the literature.  A small systematic offset (2 \kms, or 0.1 pixel)
was found between the blue and red spectra and the literature values, which
is well within the expected error for these relatively high signal-to-noise
stellar observations.  The accuracy of the spatial rectification was confirmed by
measuring the location of the each galaxy's center of light at several different
wavelengths; spatial offsets from one end of the spectra to the other were minimal
for both sides of the spectrograph.

Each galaxy observation consisted of multiple 1200 sec exposures.  
Night sky lines were removed from the two dimensional images before
the individual images were averaged together.  The final combined images were 
flux calibrated based on the sensitivity function derived from the flux 
standards observed each night.  One-dimensional (1-d) spectra of the stars and galaxies were 
extracted from the 2-d spectra using the task APALL. 
Integrated spectra for each galaxy are shown in Figure \ref{fig:spec}. 

Several different techniques were used to extract kinematic data from the observed
spectra.  The first technique employed was a straight--forward Gaussian fit to 
the strongest lines (the 3 Mg Ib lines on the blue side and the Ca triplet on 
the red side).  To improve the signal--to--noise ratio, the 2-d images were boxcar 
smoothed to a spatial resolution of 1.87\arcsec/pix for the blue camera 
(binned by 3 pixels) and 2.41\arcsec/pix for the red camera (binned by 5 pixels).  
The rotation curves were determined from the centroid of the Gaussian fits and 
velocity dispersions were determined by subtracting the instrumental
width (derived from the standard star observations) in quadrature from the
observed line widths.  While the values obtained from this simplistic measurement 
technique are not reported here, the Gaussian fitting procedure provided a baseline 
value for the rotation curves and velocity dispersion measurements described below.
  
Fourier analysis provides a more robust measure of the stellar kinematics, 
particularly when line shapes are not expected to be perfect Gaussian profiles.  The 
rotation curves for each galaxy were derived from cross--correlation analysis of 
the observed spectra with radial velocity template stars (Table \ref{tab:stars}).   
In order to determine which of the radial velocity standards provided the best
template match to the galaxy spectra, all of the radial velocity standards observed
during the same observing run were cross-correlated with the 1-d galaxy spectra
(Figure \ref{fig:spec}) using the task FXCOR in the RV package.
The radial velocity standards were then rank ordered based on
the derived Tonry-Davis Ratio \citep[TDR,][]{TD79} for each galaxy.  The later spectral types 
(G5 III - K2 III) proved to have the best correlation coefficients when cross 
correlated with the dwarf elliptical galaxy spectra; in fact, the best template 
match for both observing runs was HD 65934, a star classified as a G8 III. 
For both observing runs, the seven stars with the highest average TDR were used 
in the subsequent analysis.

Prior to further cross--correlation analysis, the 2-d images were boxcar smoothed
to a spatial resolution of approximately 2\arcsec/pix (see above), 4\arcsec/pix,
and 8\arcsec/pix to increase the signal--to--noise ratio.   Each row of
each 2-d image was cross--correlated with the selected subset of radial velocity
standards using XCOR in the contributed REDSHIFT package of STSDAS.
The task XCOR was selected for the final analysis because it returned both
a redshift and a velocity dispersion measurement.  In this analysis,
the derived centroid values were not sensitive to the spectral type of the template star;
typical individual differences between the systemic velocity measured by each of the
templates were on the order of 6 \kms~on the red side and 12 \kms~on the blue side
and relative velocity differences were much smaller than that.  The systemic heliocentric
velocities for each galaxy are tabulated in Table \ref{tab:results}.
The error estimates quoted in Table \ref{tab:results} are derived from the 
dispersion of results from the different analysis techniques and cross correlation    
of the galaxy spectra with different template stars.  These errors should be 
considered indicative of both systematic and random errors; typically, the
quoted errors are larger than the formal errors returned by the IRAF tasks.

In general, when smoothed to a spatial resolution of 2\arcsec/pix,
the galaxy rotation curves could be traced to radii of approximately
20\arcsec~or 1.56 kpc at the distance of Virgo.     There were no significant 
differences between the rotation curves derived from the Mg Ib and those derived 
from the Ca triplet lines.  The derived rotation curves are shown in Figure 
\ref{fig:rot}, with the maximum velocity width denoted by dashed lines.  Except
for VCC 1075 and VCC 1857, all of the plots in Figure \ref{fig:rot} are from
the data binned to 2\arcsec~resolution.   Seven of the sixteen galaxies have 
clear evidence of a velocity gradient across the stellar component; the 
rotational component is easily discernible in the raw image for all of these 
galaxies.  The slope of the rotation curve was derived from linear regression
analysis of all of the data points (both blue and red cameras) and
is tabulated in Table \ref{tab:results}.  The representative maximum rotation
velocity (V$_{\rm rot}$) was derived from the fitted rotation curve and the
maximum radius at which the rotation curve could be measured.    The rotation
velocities are tabulated in Table \ref{tab:results}.  Five of the galaxies in
the present sample are also included in \citet{GGvM03}; in all cases, the
present rotation curves trace the galaxy kinematics to a significantly larger
radius due to both the limited slit length of the Keck observations and to
signal--to--noise ratio concerns.  Thus, the derived maximum rotation
velocities are larger than previously reported. 

The same cross-correlation analysis yielded a velocity dispersion measurement for
each galaxy.  The velocity dispersion measurements were calibrated by analyzing the
artificially broadened spectra of one of the template stars.  Based on the stellar
template, observations with the blue camera did not have sufficient resolution to
provide accurate velocity dispersion measurements of these galaxies,
while the Ca line observations had sufficient resolution to measure velocity 
dispersions greater than 35 \kms.  The measured velocity dispersions are shown 
in Figure \ref{fig:rot} and the central ($\sigma_0$) and median ($\sigma_m$)
 values are tabulated in Table \ref{tab:results}.  The median velocity dispersion
is derived from the average of the measurements at all radii.
The typical velocity dispersion for the galaxies in this sample is 44 $\pm$ 5 \kms. 
In general, given the relatively poor spectral resolution of both cameras
and low velocity dispersions of the galaxies, the reported velocity dispersions
should be considered indicative upper limits.

For comparison, three of these galaxies have velocity dispersions in the published
literature [VCC 543 and VCC 1122, \citet{BBF92} and \citet{PGCSBG02}; VCC 917, 
\citet{GGvM02}].  The mean velocity dispersion for VCC 1122 reported in \citet{BBF92}
is larger than the value reported here, but is consistent with the central
velocity dispersion measurement.  Figure \ref{fig:comp} shows a direct comparison 
of the line shape expected for a template star broadened by the velocity dispersion 
for the integrated spectrum of VCC 1122; the derived velocity dispersion is 
clearly a good match to the observed spectra.  The velocity dispersion of 
VCC 543 reported here is comparable to that of \citet{BBF92}; similar good 
agreement is found between the new observations of VCC 917 and those reported 
in \citet{GGvM02}, although the
reported systemic velocity  of VCC 917 is significantly different than that reported here.

\section{Rotation in Dwarf Elliptical Galaxies}

The derived rotation velocities and velocity dispersions indicate that several 
dwarf elliptical galaxies in the Virgo cluster have significant rotation components.
The ratio of the rotation velocity to velocity dispersion for the Virgo
dwarf elliptical galaxies is shown in Figure \ref{fig:anisotrop} and tabulated in
Table \ref{tab:results}.  
Note, however, that the anisotropy parameters, (v/$\sigma_m$)*,
listed in Table \ref{tab:results} and shown in Figure \ref{fig:anisotrop} should
be considered representative lower limits for the dEs.   Not only are the derived velocity
dispersions representative upper limits, the present observations only sampled the
kinematics along the major axis; if the dEs have a significant velocity gradient
along a different axis, such as found in some dI galaxies 
[e.g., NGC 625, \citep{CCF00} and NGC 5253, \citep{KS95}], the
derived anisotropy parameters will be underestimates of their true values. 
Also shown in Figure \ref{fig:anisotrop} are the anisotropy
parameters for giant elliptical galaxies from \citet{BBF92} 
and the dwarf elliptical galaxies in \citet{PGCSBG02}, \citet{GGvM02}, \citet{SP02},
and \citet{DDZH01,DDZH03}.
Five of the dwarf elliptical galaxies in the Virgo sample have a significant
rotation component.  These galaxies overlap the locus of anisotropy 
parameters found in giant elliptical galaxies, where the lower luminosity 
ellipticals tend to have large anisotropy parameters.  
Eight of the remaining dwarf elliptical galaxies have
modest anisotropy parameters, while only one, VCC 1261, has an extremely low
value.  These results are significantly different than those reported by
\citet{GGvM02} and the early studies of \citet{BN90} and \citet{BPN91} 
where it appeared that dwarf elliptical galaxies had no significant rotation component.
However, the recent observations by \citet{PGCSBG02} and \citet{GGvM03} revealed a handful of 
dwarf elliptical galaxies with significant rotation. Thus, it appears that the kinematics of some 
dwarf elliptical galaxies are dominated by rotation, not random motions.

While this result is counter to the results reported in earlier studies,
it should not be surprising.  The dwarf elliptical galaxies in this sample are all 
located in the region of strong overdensity that characterizes the Virgo Cluster.  It is well known
that galaxies located near the cluster center are deficient in neutral
hydrogen relative to their counterparts in the field \citep[e.g.,][]{GH83,SMGGGH01}.
 For giant spiral galaxies, the loss of their ISM may
result in modest morphological changes, including suppression of their star
formation activity \citep[e.g.,][]{KK98}.  However, despite these changes,
gas-deficient giant spiral galaxies are likely still to be classified as 
spiral galaxies since there are striking morphological differences between
giant spiral and elliptical galaxies.  This is not the case for dwarf galaxies.  Dwarf
elliptical galaxies and dwarf irregular galaxies share common morphological
parameters; both have approximately exponential surface brightness profiles
and have similar central surface brightnesses and scale lengths (see Section 4 for
further discussion of the similarities between dEs and dIs). 
 Thus, their structural parameters do not provide
a strong morphological distinction between dwarf elliptical galaxies and
 dwarf irregular galaxies; rather,  their morphological classification is
highly dependent on the patchy appearance (or absence thereof) due to
irregular star formation activity across the surface of the galaxy. 
If star formation ceases in a dwarf galaxy, the resulting system is likely
to be classified as a dwarf elliptical galaxy once the massive stars, which
provide the patchy contrast, have evolved off of the main sequence.
 In fact, it is this morphological commonality that first led to the idea that dwarf
elliptical galaxies could evolve passively from dwarf irregular galaxies
after a starburst episode \citep[e.g.,][]{LF83,K85,LT86,DH91,J94,PLFT96,SHRCK98}.

If the dwarf elliptical galaxies in this sample are the end products of
dwarf irregular galaxies which have been stripped of their ISM, the observed
rotation velocities should be similar to the rotation velocities of normal
dwarf irregular galaxies.   The correlation between rotation width  (corrected
for inclination) and absolute magnitude for the dwarf elliptical galaxies is 
shown in Figure \ref{fig:tf}.  Also shown in Figure \ref{fig:tf} are the values
for dwarf irregular galaxies compiled in \citet{vZ01} 
and the correlation derived for spiral galaxies \citep{TP00}.
In general, dwarf irregular galaxies have a large scatter in the Tully-Fisher diagram, 
in part due to uncertain inclination and velocity dispersion corrections.  However, 
there is still a general trend that more luminous dwarf irregular galaxies have
larger rotation widths.  Many of the  dwarf elliptical galaxies also appear to 
follow this trend;  the exceptions are VCC 917, 1122, 1261, 1308, 1514, and 2050,
which are under-rotating for their observed luminosity.   These galaxies all have
kinematic slopes which are less than 20 \kms~kpc$^{-1}$; we designate these galaxies
as the ``non-rotating'' sample.  Those that follow the Tully-Fisher relation (VCC 178, 437, 
543, 965, 990, 1036, 2019) are designated the ``rotationally supported'' sample.
Note that VCC 1743 and VCC 1857 also appear to be rotationally supported, but
their observations are too poor quality to include in the statistics.  These
groupings are similar to those seen in the anisotropy diagram (Figure \ref{fig:anisotrop}),
with the exception that VCC 178 and 965 fall in the ``modest'' (v/$\sigma_m$)* grouping.
We elect to separate the kinematic samples based on the amplitude of the slope of
the rotation curve and location in the Tully-Fisher diagram since the velocity 
dispersion measurements are not as secure as the rotation curve measurements for
these observations.

Note that Figure \ref{fig:tf} compares stellar (dE) and
gaseous (dI and spirals) rotation curves.  While the stars and gas are expected to
be kinematically coupled in normal galaxies, it is necessary to consider whether
the stellar rotation curves sample fully the rotation velocities of these galaxies.
As mentioned above, the observations were limited to the optical major axis;
for most galaxies, this is the axis with the largest velocity gradient, and
thus it is reasonable to assume that the rotation curves measured here are
representative.  However, the stellar rotation curves were traced out
to only a modest radius whereas the neutral gas velocity widths include emission from
gas that extends well beyond the optical galaxy.  Nonetheless, if the dwarf elliptical galaxies
and dwarf irregular galaxies have similar kinematic properties, it should be
possible to correct for this effect on a statistical basis.  

For comparison, we consider
the neutral gas distribution and kinematics for a sample of 20 galaxies selected
from the optical imaging samples of \citet{vHS97} and \citet{vZ01}, i.e., a subset
of the dwarf irregular galaxies shown in Figure \ref{fig:tf}.  The resolved HI
distribution and kinematics were obtained with Very Large Array during several 
observing runs from 1993 - 2000.  While the majority of the HI synthesis data
is as yet unpublished, the dI sample includes the galaxies in \citet{vHSB97} with
distances scaled to an H$_0$ of 75 \kms~Mpc$^{-1}$.
A histogram of the extent of the HI distribution to optical 
scale length for the dI sample is shown in Figure \ref{fig:slope}a.  
To determine the statistical correction to convert from a stellar
rotation width to a neutral gas rotation width, we measured the ratio of the
maximum rotation velocity to the slope of the rotation curve.
A histogram of the V$_{\rm max}$/slope for dwarf irregular galaxies is
shown in Figure \ref{fig:slope}b, where the slope of the rotation curve is in units
of \kms~per scale length.  The mean value is 2.45 $\pm$ 0.58 scale lengths, 
which shows remarkable (and perhaps co-incidental)
agreement with the use of the velocity at 2.2 scale lengths as a fiducial measure of spiral galaxy
rotation curves \citep[e.g.,][]{CR99}.
Based on these results, it is likely that the stellar rotation curves obtained here underestimate
 the maximum rotation velocity for the dwarf elliptical galaxies since most were traced out
to only 1 - 1.5 scale lengths.  
Arrows in Figure \ref{fig:tf} indicate the location
of each galaxy if the stellar rotation curves were traced to 2.45 scale lengths and
the galaxy dynamics follow similar trends as those of the neutral gas in  dI galaxies.

Given the uncertainties associated with this velocity width correction, it is premature
to draw far reaching conclusions from the revised velocity widths.  Nonetheless,
it is important to note that such measurements could provide significant constraints
on the luminosity evolution of dE galaxies.  In particular, these rough correction 
factors indicate that the progenitor galaxies were at most only 2 magnitudes more
luminous than the present population (i.e., the revised rotation widths are comparable
to those of galaxies 2 magnitudes more luminous).   If these results are confirmed
by observations that fully trace the galaxy dynamics, the
progenitor population of Virgo dEs must be restricted to moderate luminosity systems.
In particular, it is unlikely that dEs evolve from significantly
more luminous galaxies.

\section{Dwarf Galaxy Evolution Scenarios}

One of the perplexing issues in galaxy evolution is the remarkable commonality 
between dwarf elliptical galaxies and dwarf irregular galaxies.  These low
mass systems have similar stellar distributions, both in terms of functional
form (exponential) and in terms of typical central surface brightnesses and
scale lengths \citep{LF83}.  Furthermore, both dIs and dEs follow the same luminosity-metallicity
relation, where the more luminous galaxies are more metal rich, regardless of
their gas content \citep{SKH,RM95}.   In fact, aside from 
their current gas content and star formation activity, the low mass galaxy classes 
are remarkably similar in stellar content and morphology. 

As mentioned in the introduction, the apparent morphological similarity between the
low mass galaxy classes led to several possible evolutionary scenarios that linked
dwarf irregular galaxies and dwarf elliptical galaxies.
However, the apparent kinematic mis-match between dEs and dIs posed a severe constraint 
for evolutionary pathways between dIs and dEs since angular momentum must be 
conserved \citep[see, e.g.,][]{BPN91,SB95,vSS01}.
However, the early studies included only a small number of dEs.
The new observations presented here indicate a
remarkable similarity between the kinematic properties of 
many dwarf elliptical and dwarf irregular galaxies; these results 
re-open the possibility that some dEs are formed from passive evolution of dIs\footnote{In
this context, passive evolution is taken to mean simple evolution of the stellar 
population without a major merger or other catastrophic event.}. 

To some degree, the question of whether dwarf elliptical galaxies evolve from
dwarf irregular galaxies is tautological since every dE {\sl must} have been
gas-rich and star-forming at some point in the past.  Standard morphological
classification would identify these gas-rich, low mass galaxies as dIs.
Thus, the relevant question is not whether a dE evolves from a dI, but what caused
the progenitors of dEs to stop forming stars and to lose their ISM. 

One popular scenario is that the progenitors were too low mass to retain their
ISM after a star formation episode \citep[e.g.][]{DS86,DG90}.  In this
scenario, the kinetic energy from supernovae explosions is sufficient
to sweep out the remaining ISM after a starburst episode.  However, more
recent calculations indicate that it may be more difficult to remove the
ISM of a dark matter dominated galaxy than previously thought \citep[e.g.,][]{DH94,ST98,MF99,FT00}.
Furthermore, detailed star formation histories of nearby dwarf Spheroidal galaxies
indicate these extremely low mass galaxies were able to retain their ISM through several star formation
episodes \citep[e.g., Carina dSph,][]{SHSHL94}.  Thus, internal gas-loss mechanisms
face several challenges to explain the diversity of stellar populations,
the relative chemical enrichment, and the relative gas-richness of low mass galaxies.

Alternatively, the key to the formation of dwarf elliptical galaxies may be
an external process.  Dwarf elliptical galaxies are the most strongly
clustered galaxy type \citep{FS89}.  They are predominantly found in high density
regions, either in galaxy clusters or located near more massive spiral and elliptical
galaxies.  Given the strong morphology-density relation for dwarf elliptical galaxies, 
it is reasonable to assume that the environment may play a crucial role in their evolution.
Segregation within the dwarf ellipticals in Virgo itself has already
been noted by several authors \citep{FS89,Oh00}, who 
point out that the nucleated dE's occupy the cluster core,
whereas the majority of non-nucleated ones are found in the peripheral
regions.
Furthermore, as discussed in Section 3, ram pressure stripping is an
efficient mechanism to remove the ISM from cluster members, regardless of total mass.
Infall of galaxies into
the Virgo Cluster is an ongoing process  \citep{TS84,COGW03}; 
one thus expects that small groups of galaxies similar to the Local Group
continue to fall into the Virgo Cluster.  Given the simple surface
mass density criteria for ram pressure stripping  \citep{GG72},
essentially all low mass dwarf irregular galaxies falling into the
Virgo cluster will be stripped of their gas \citep[e.g.,][]{MBD03}.  
Indeed, there is evidence
of a very recent case of stripping of a Virgo dwarf galaxy 
\citep[UGC 7636,][]{STE87,PT92}.  Unless
one imagines that all dwarf galaxies that fall into the Virgo cluster
are already dEs (and the fact that they are falling in from low
density environments argues strongly against this), then it would
appear to be inevitable that some of the dEs in the Virgo cluster
are stripped dIs.

Since dwarf irregular galaxies are dark matter dominated, removal of their 
ISM will have only a modest effect on their kinematics.  However,
subsequent passages through the cluster could further disrupt the
stellar kinematics via galaxy harassment or merging events.  Thus, if 
dIs are converted into dEs via ram pressure stripping, 
one might expect a correlation between the location of dwarf 
elliptical galaxies with significant rotation and the cluster center.
While the present sample is relatively small, Figure \ref{fig:cluster} 
shows a hint of such a relationship, in the sense that galaxies with 
little-to-no rotation appear
to be located predominantly in the cluster core or in high density
clumps, while those with significant
rotation are in the outskirts of the cluster.  This is consistent with the
idea that gas-rich dwarf galaxies are stripped of their gas during a passage
through the intracluster medium; the non-rotating
dwarf elliptical galaxies may be remnants of gas-rich systems that have
made multiple passes through the ICM, or had a catastrophic event occur.  
We caution, however, that the present sample size is relatively small and 
biased toward the more luminous cluster dEs; further observations of a 
larger sample of dwarf elliptical
galaxies throughout the cluster environment are needed to investigate 
this suggestive trend.

Although we have argued above that the available observations do not
allow us to establish the relative strengths of the rotationally
supported and non-rotationally supported dE populations, we cannot
close without some speculation on the presence of non-rotating
dE galaxies.  Following the same line of argument given above, i.e., that
it is almost inescapable that some of the dEs observed in the Virgo
cluster must have lost their gas to ram pressure stripping, it is
also likely that some dwarf galaxies entered the Virgo cluster
as dEs before they encountered sufficient hot gas for stripping.
It also seems inescapable that some dI galaxies interact
tidally with more massive galaxies, and thereby lose gas in this manner.
Given that it is difficult to quantify the relative
importance of these three different channels of cluster dE galaxy
formation, we caution against single channel evolutionary
scenarios.  Specifically, we note that explanations for the
density--morphology relationship in the Local Group may be
irrelevant for the cluster density--morphology relationship.
A more comprehensive study of kinematics in dE galaxies in
both environments is warranted.   Based on the data in hand,
however, it is clear that environmental pathways for dI to
dE conversion must be considered; if further observations
support the trend indicated in Figure \ref{fig:cluster},
relevant theories for the formation and evolution of dwarf galaxies must
include environmental factors as well as internal processes.

\section{Conclusions}

We present long-slit optical spectroscopy along the major axis of 16 dwarf elliptical
galaxies in the Virgo Cluster.  The major results of these observations are as follows:

(1) Approximately half of the Virgo dE sample has a significant rotation component, with
rotation curve slopes $>$ 20 \kms kpc$^{-1}$.  

(2) Five of the 14 galaxies with measured velocity dispersions have anisotropy 
parameters (v/$\sigma_m$)* $>$ 1., indicating significant rotational flattening.  
Only one of the remaining 9 dEs has (v/$\sigma_m$)* $<$ 0.3.  The remainder have
modest anisotropy parameters, indicating that rotational flattening may be
significant for the majority of dwarf elliptical galaxies in this sample.

(3) Based on the observed maximum rotation velocities, the rotating dwarf
galaxies appear to follow the Tully-Fisher relation for gas-rich dwarf and spiral
galaxies.

These kinematic results re-open the possibility that dwarf elliptical galaxies
may be the end products of dwarf irregular galaxies that have lost their
ISM via ram pressure stripping or other non-catastrophic processes.  The
morphological similarities between gas-rich dwarf irregular galaxies and 
gas-poor dwarf elliptical galaxies appear to extend beyond the stellar 
distributions and metal enrichment; many gas-rich and gas-poor dwarf
galaxies have common dynamical properties, as would be expected
for passive evolution models.

\acknowledgments
We dedicate this paper in memory of J.\ Bev Oke, whose spectrographs
have enabled a plethora of science for many decades and whose
dedication and kindness encouraged several generations of scientists
to use those spectrographs well.
LvZ thanks Elizabeth Barton for many thought-provoking conversations about
galaxy formation and evolution.
This research has made use of the NASA/IPAC Extragalactic Database (NED)
which is operated by the Jet Propulsion Laboratory, California Institute
of Technology, under contract with the National Aeronautics and Space
Administration.  LvZ acknowledges partial support from the 
Herzberg Institute of Astrophysics and the National Research Council of Canada; 
LvZ also acknowledges partial support from Indiana University.
EDS is grateful for partial support from a NASA LTSARP grant No. NAG5-9221
and the University of Minnesota. MPH has been supported by NSF grants AST-9900695
and AST-0307396.

\vfill
\eject

\begin{table}
\dummytable\label{tab:global}
%this table contains the list of global parameters
\end{table}

\begin{table}
\dummytable\label{tab:obs}
%this table contains the list of observations
\end{table}

\begin{table}
\dummytable\label{tab:stars}
%this table contains the list radial velocity standards
\end{table}

\begin{table}
\dummytable\label{tab:results}
%this table contains the systemic velocities, rotation velocities, velocity dispersions
\end{table}

\psfig{figure=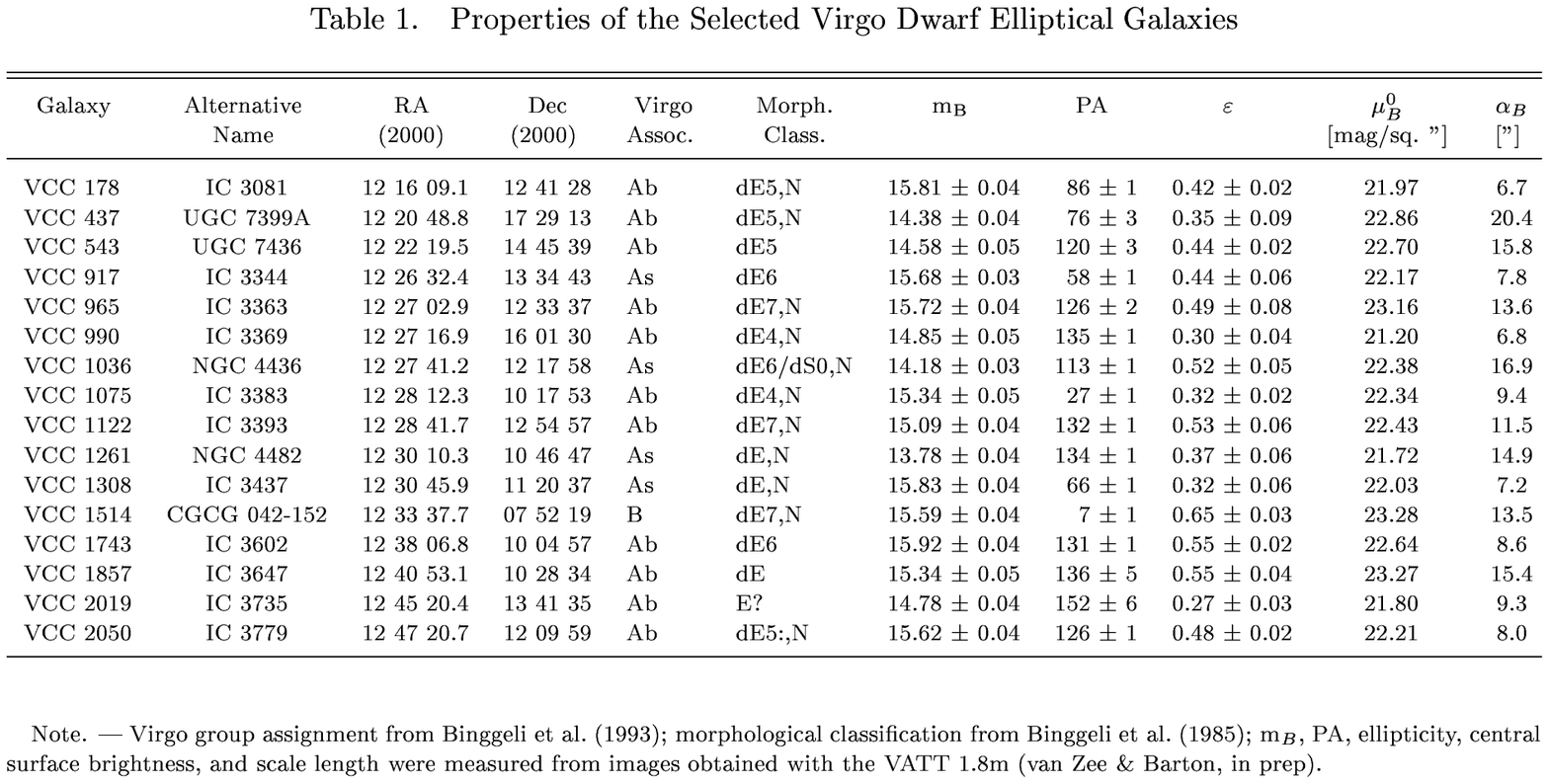,height=25.cm}

\psfig{figure=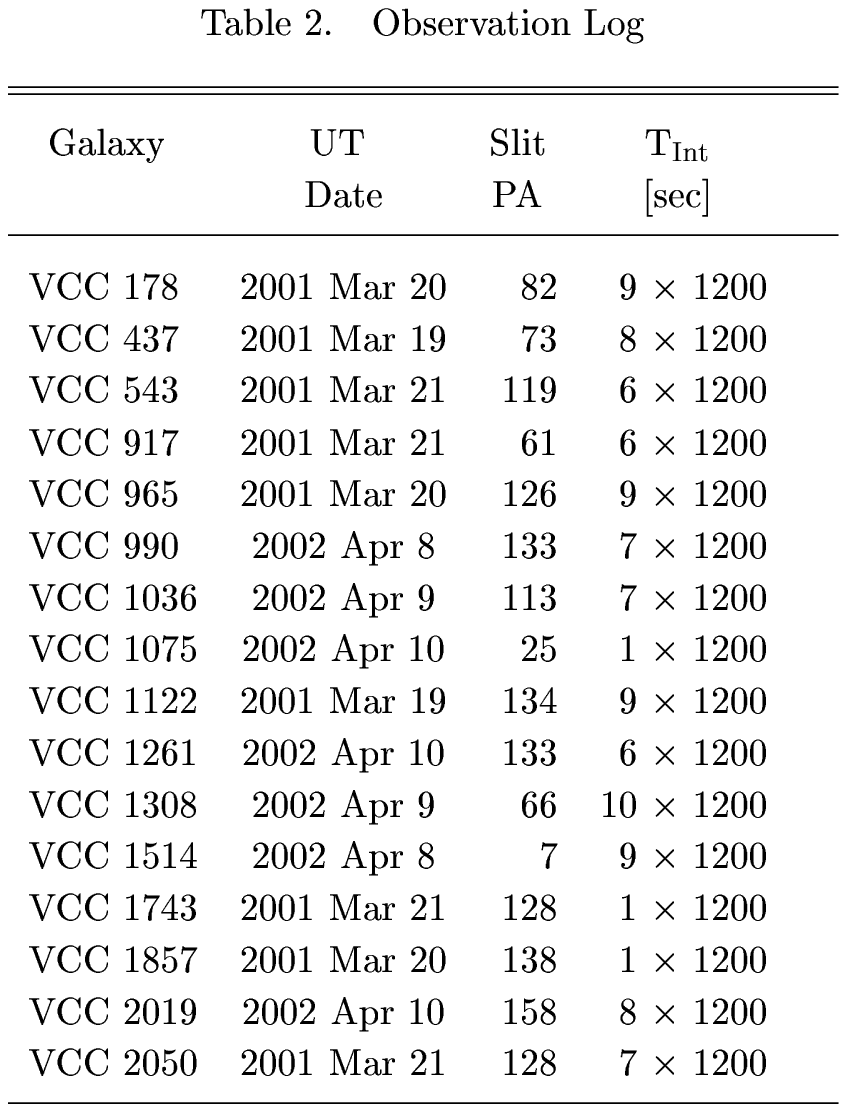}

\psfig{figure=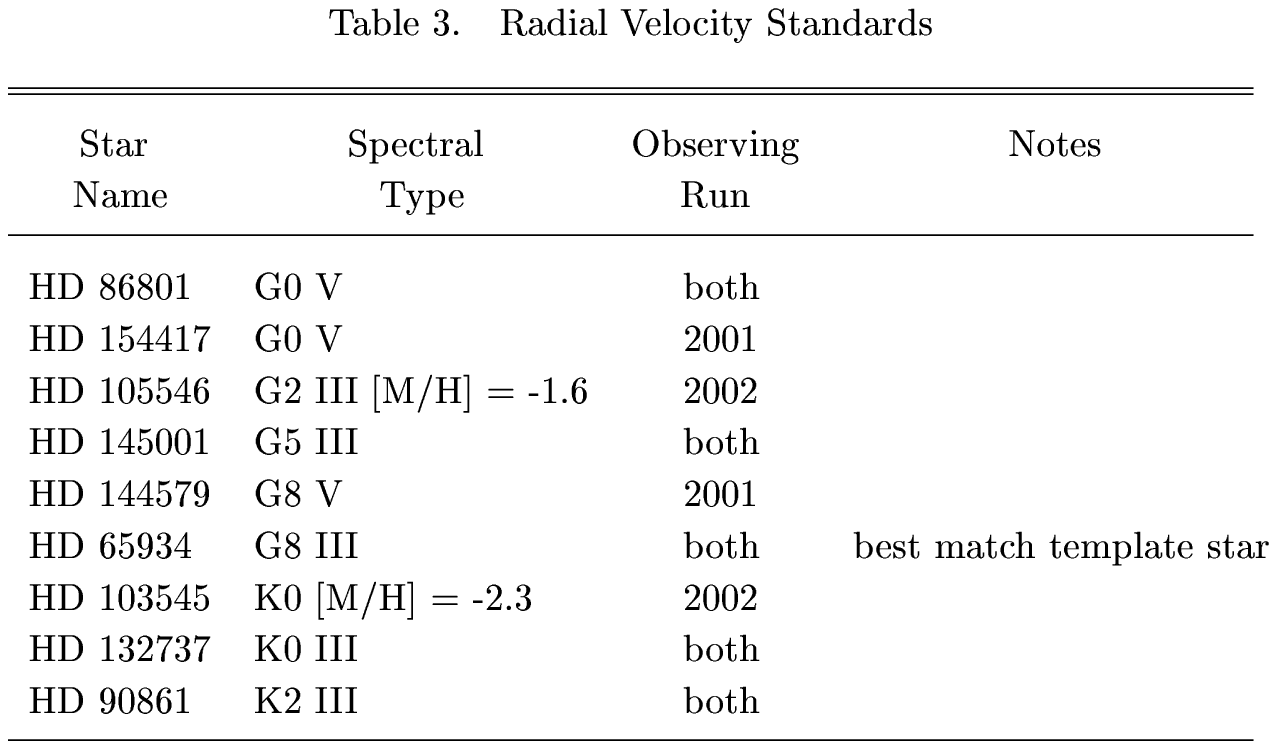}

\psfig{figure=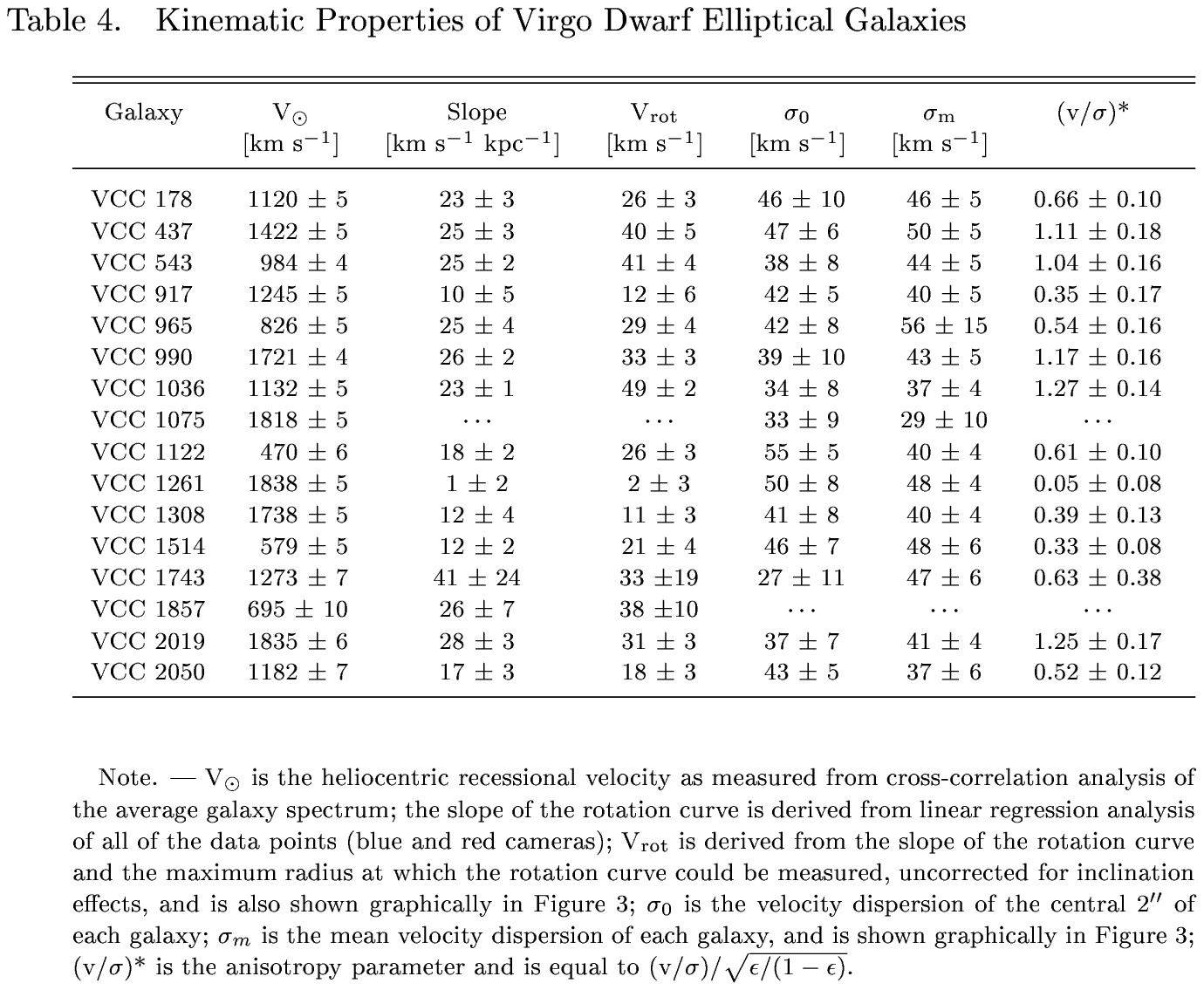}

\psfig{figure=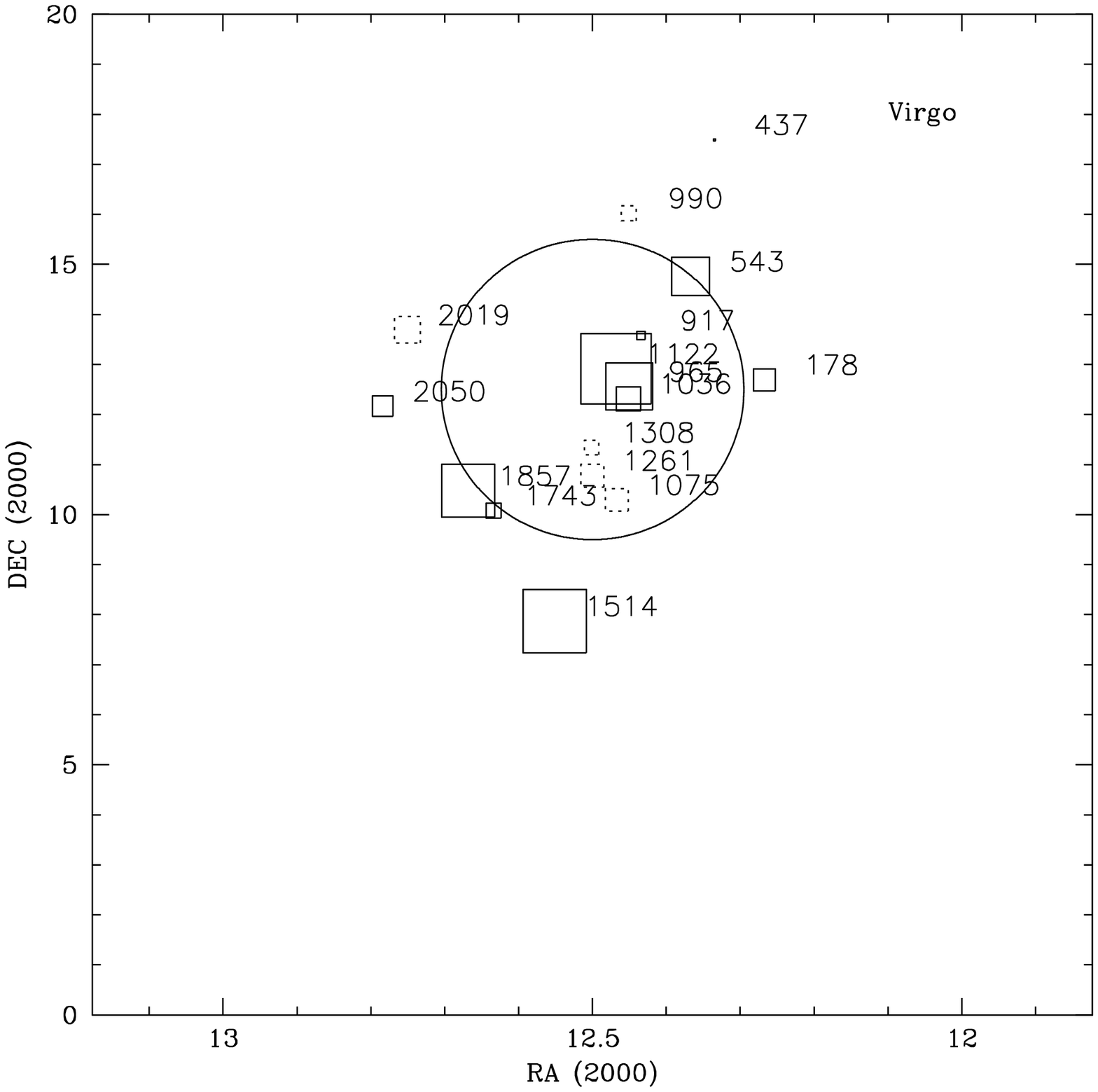,height=15.cm}
\figcaption[vanzee.fig1.ps]{Location of the dwarf elliptical galaxies in the Virgo
 Cluster.  The size of the box is indicative of the recessional velocity relative
to 1050 \kms, with solid squares indicating galaxies that are blueshifted and 
dotted squares indicating galaxies that are redshifted relative to this value. 
The circle denotes the inner 3\arcdeg~radius of the Virgo Cluster.
The sample includes dwarf elliptical galaxies in a range of environments 
within the Virgo Cluster.
\label{fig:virgo} }

\psfig{figure=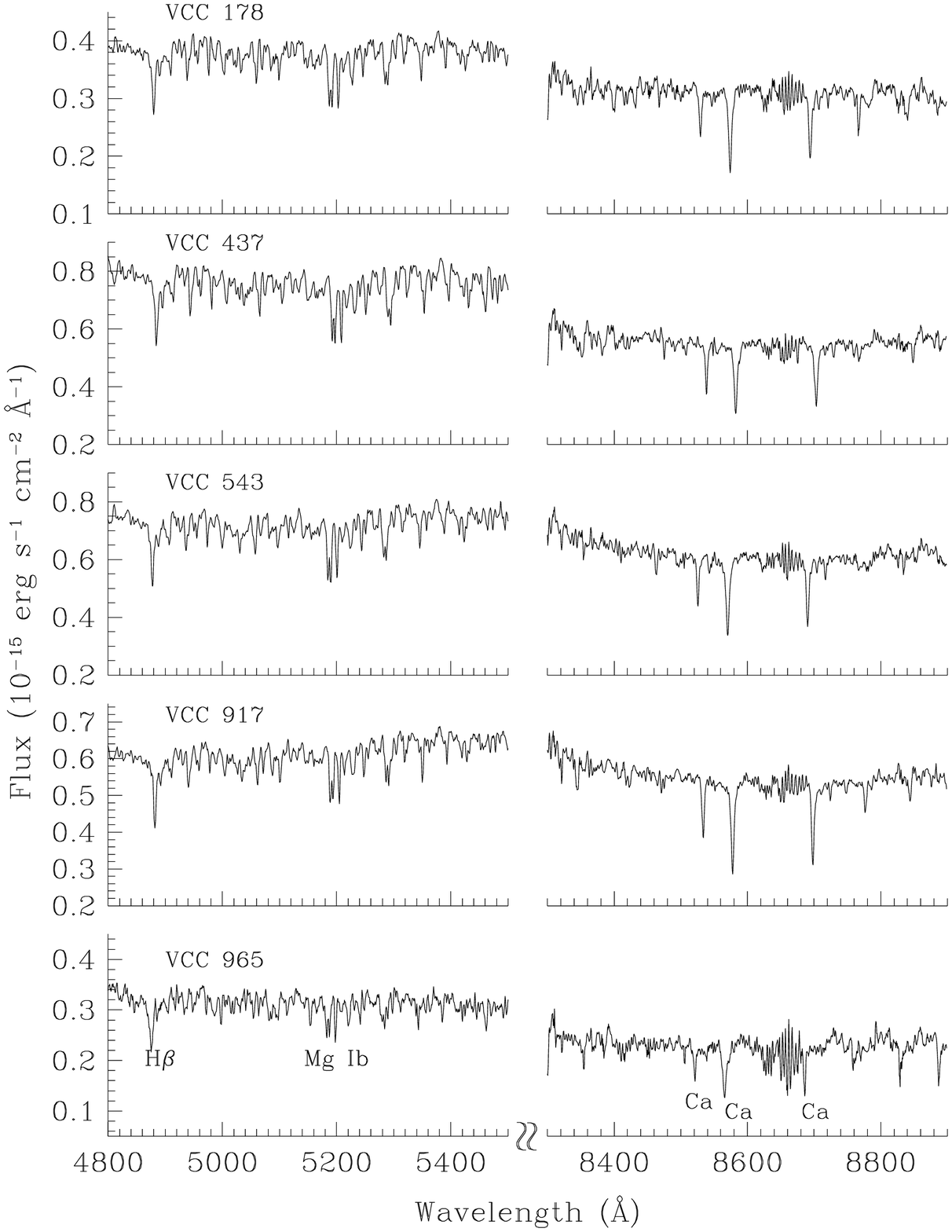,height=15.cm}
\figcaption[vanzee.fig2.ps]{Optical spectra of sixteen dwarf elliptical galaxies
summed over the entire spatial extent.  The Mg Ib and Ca triplet lines were
used to trace the stellar kinematics of each galaxy. \label{fig:spec} }

\psfig{figure=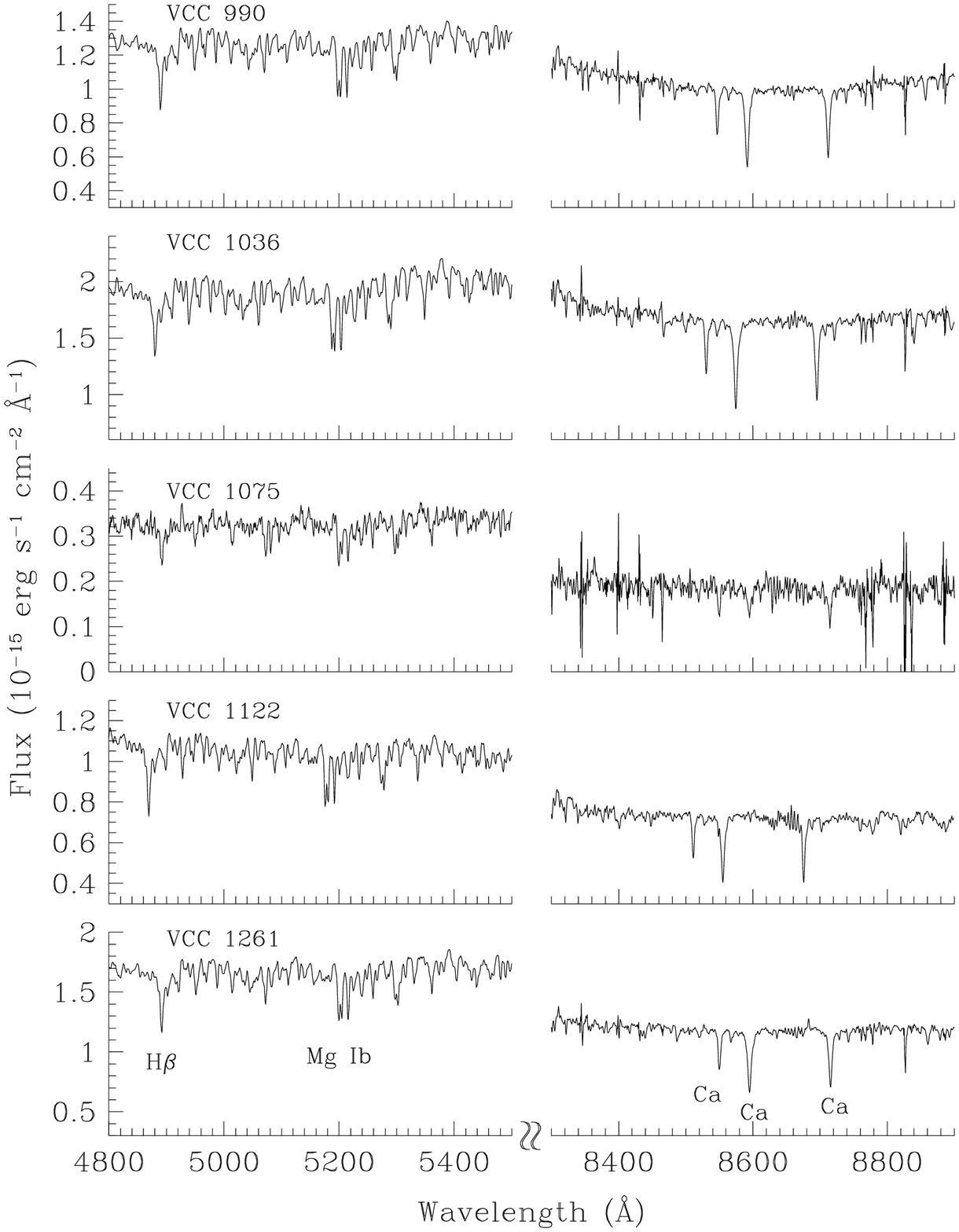,height=15.cm}

\psfig{figure=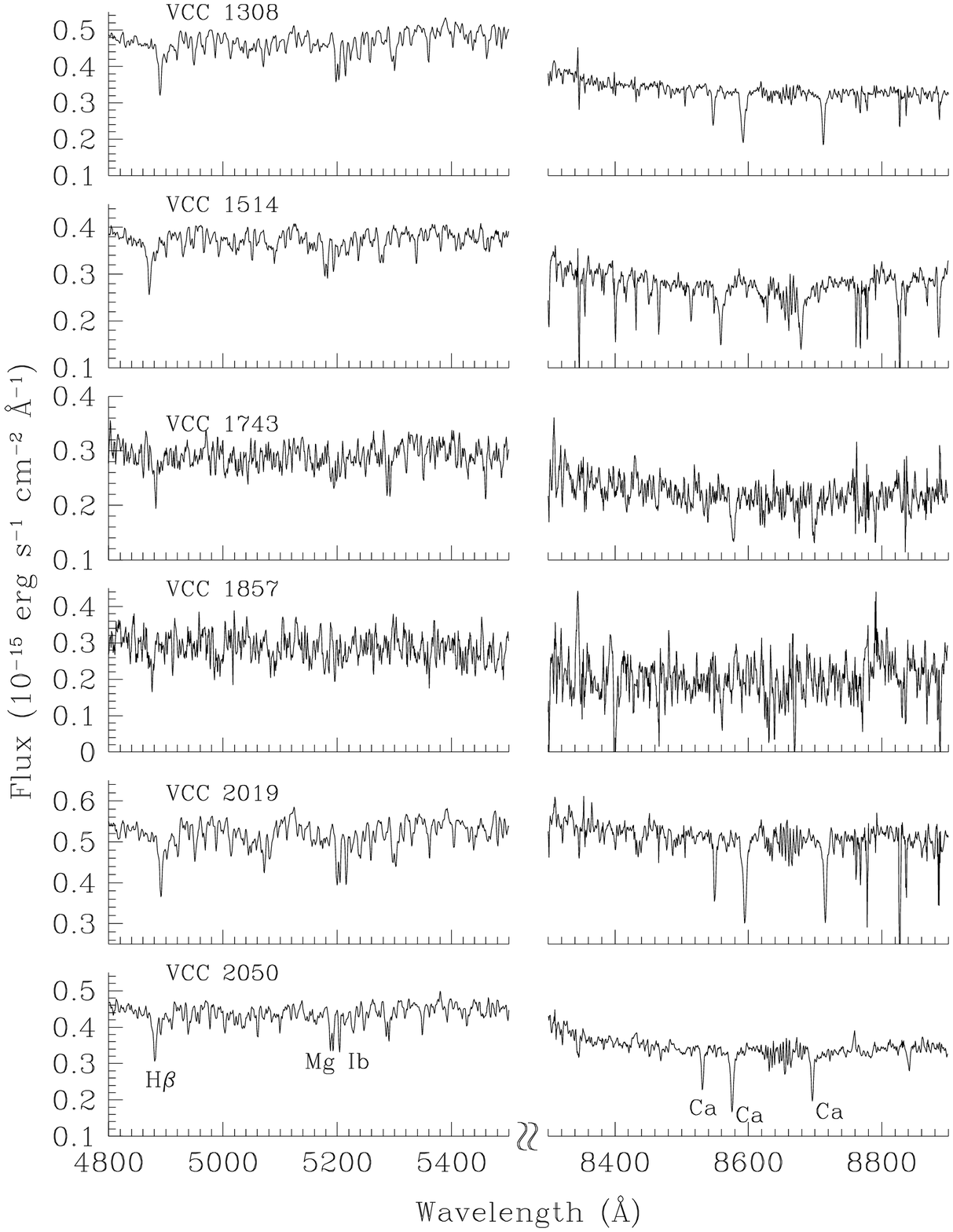,height=15.cm}

\psfig{figure=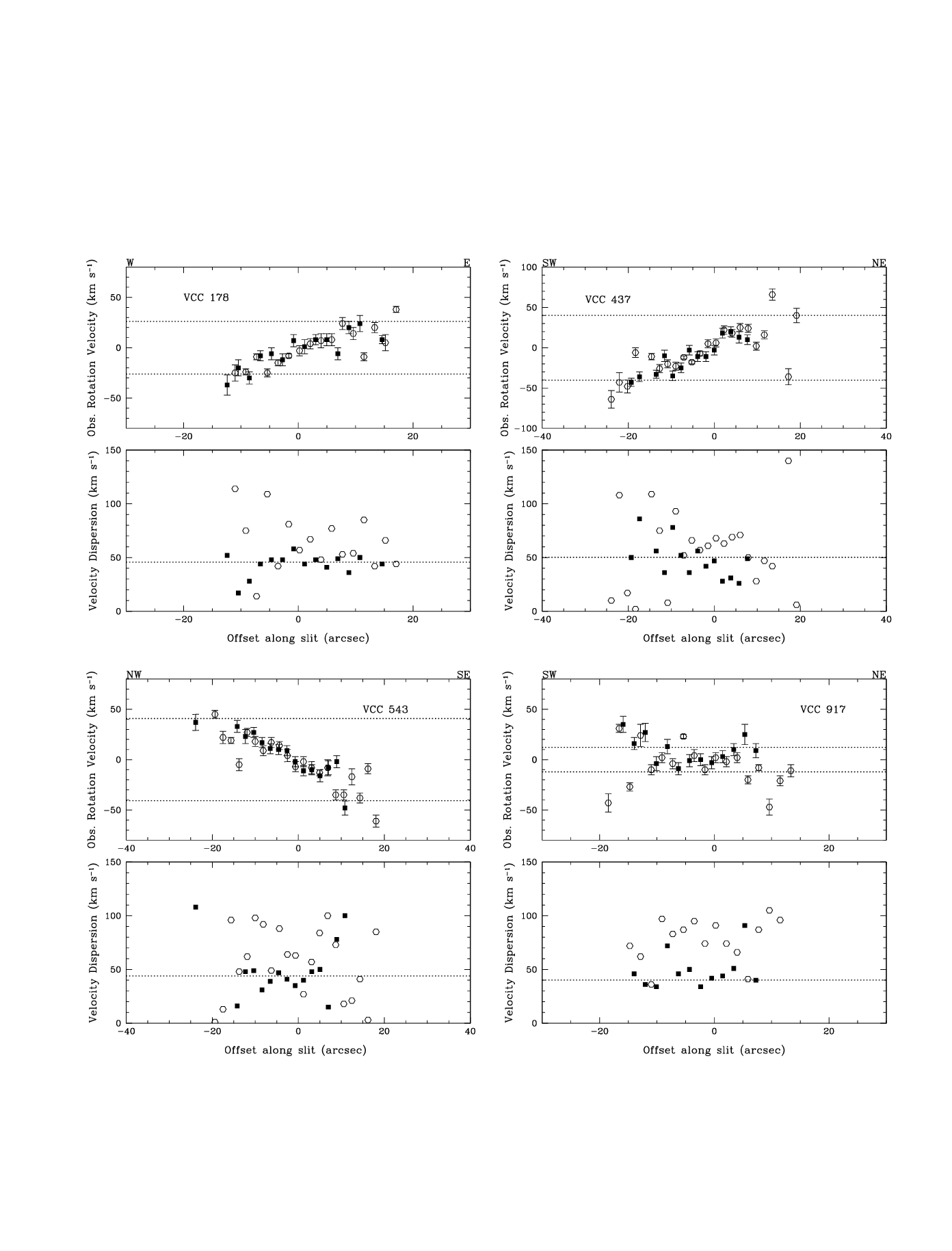,height=20.cm}
\vskip -1.3truein
\figcaption[vanzee.fig3.ps]{Rotation curves of sixteen dwarf elliptical galaxies.
The top panel shows the rotation curve derived from the Mg Ib (open circles) and 
Ca triplet (filled squares) lines.  The dashed lines indicate the maximum velocity 
width measured from the observed rotation curve.  The bottom panel shows
the velocity dispersion as a function of radius as derived from the Mg Ib and 
Ca triplet lines.   In most cases, the Mg Ib measurements are upper limits to 
the velocity dispersion; the dashed lines denote the mean value from the Ca triplet
measurements.  Seven of the sixteen dwarf galaxies have a significant 
velocity gradient across their stellar distribution.
\label{fig:rot}}

\psfig{figure=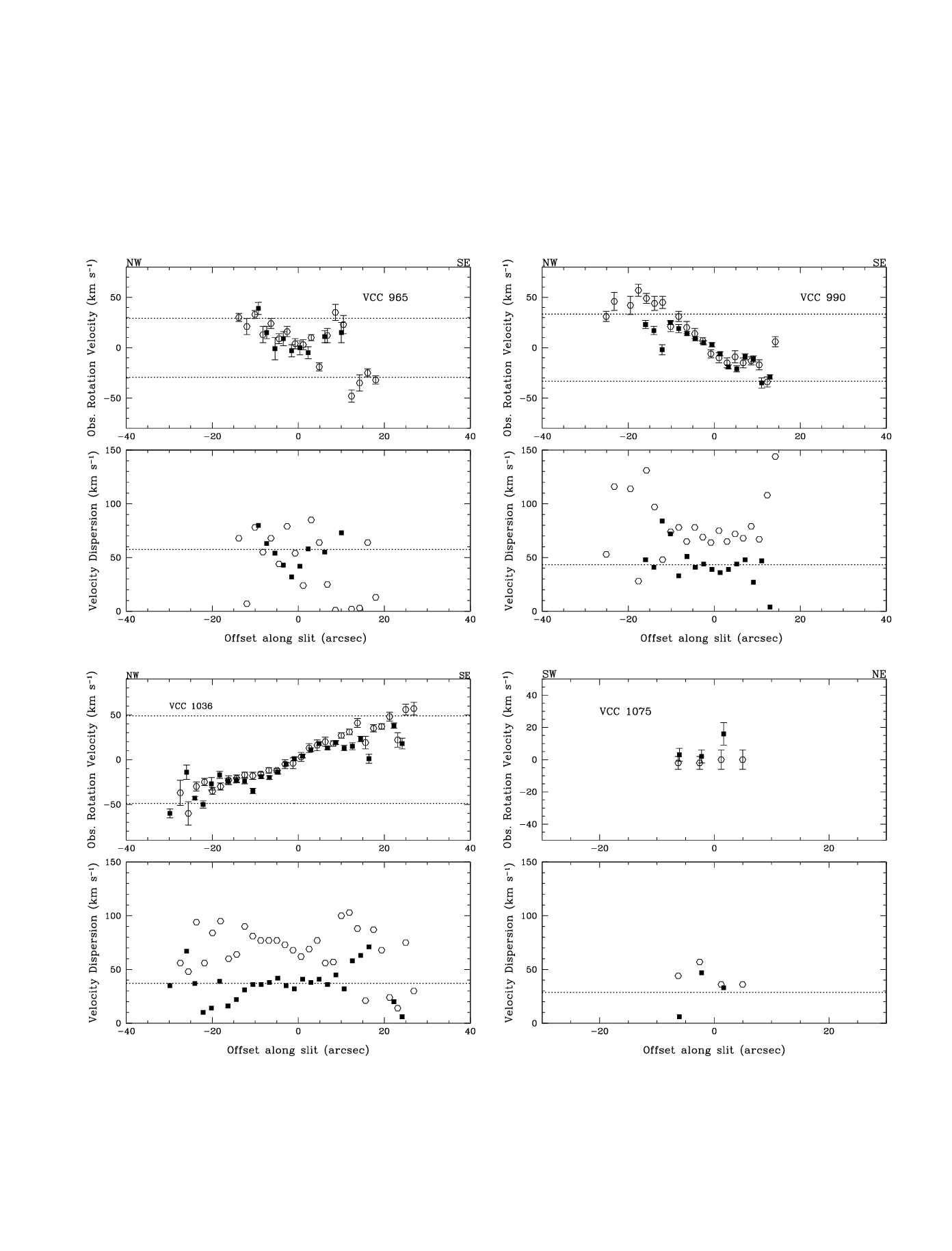,height=22.cm}

\psfig{figure=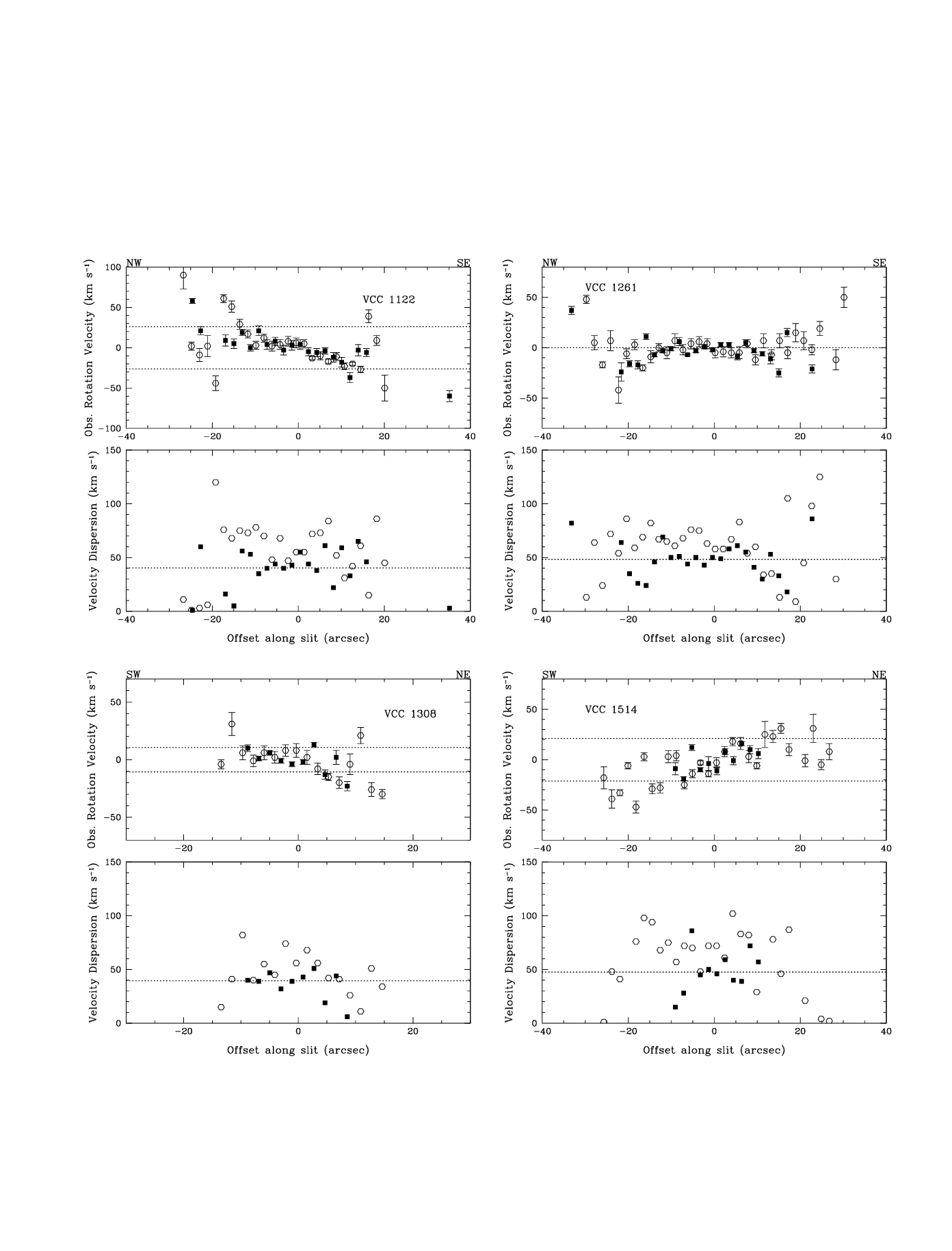,height=22.cm}

\psfig{figure=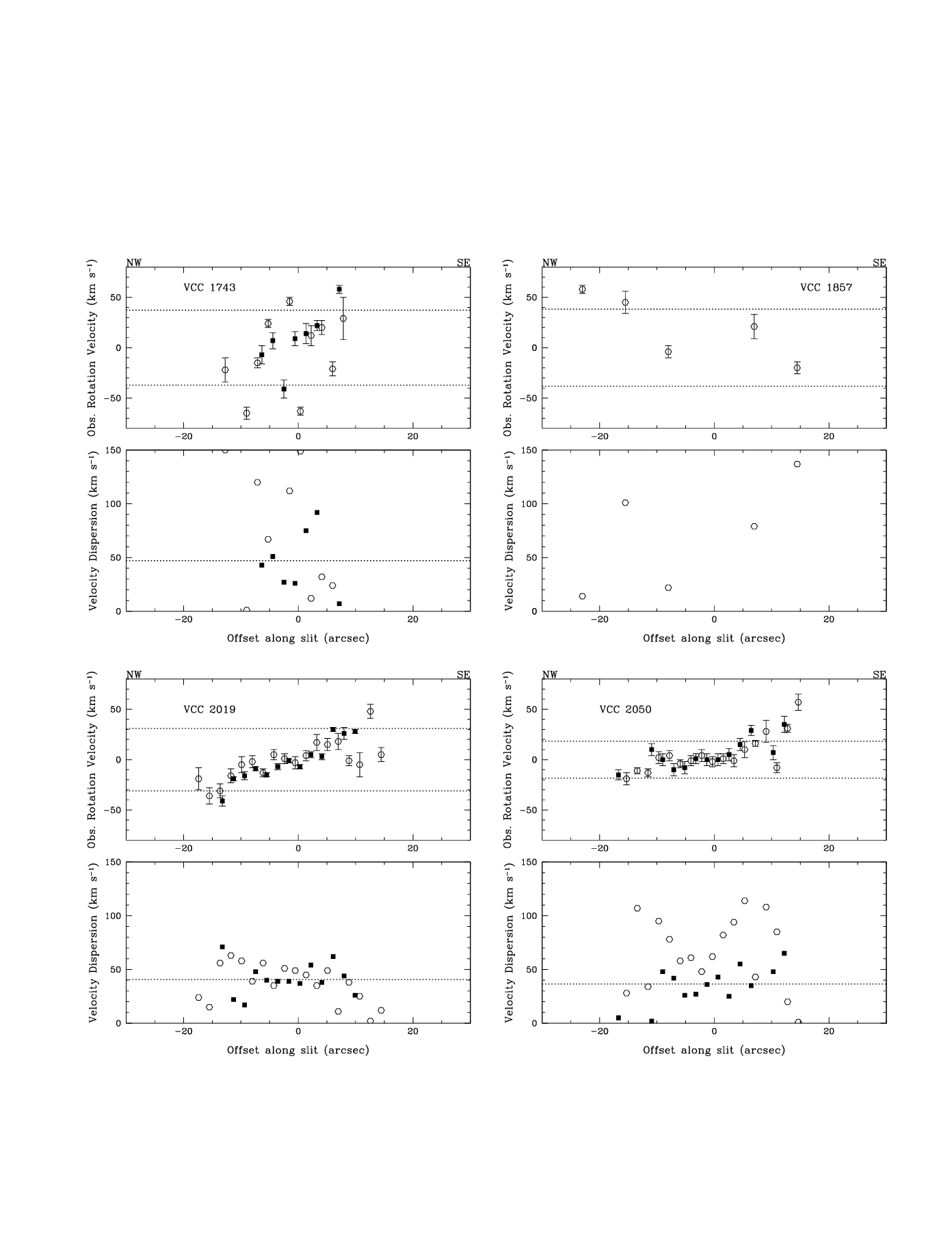,height=22.cm}

\psfig{figure=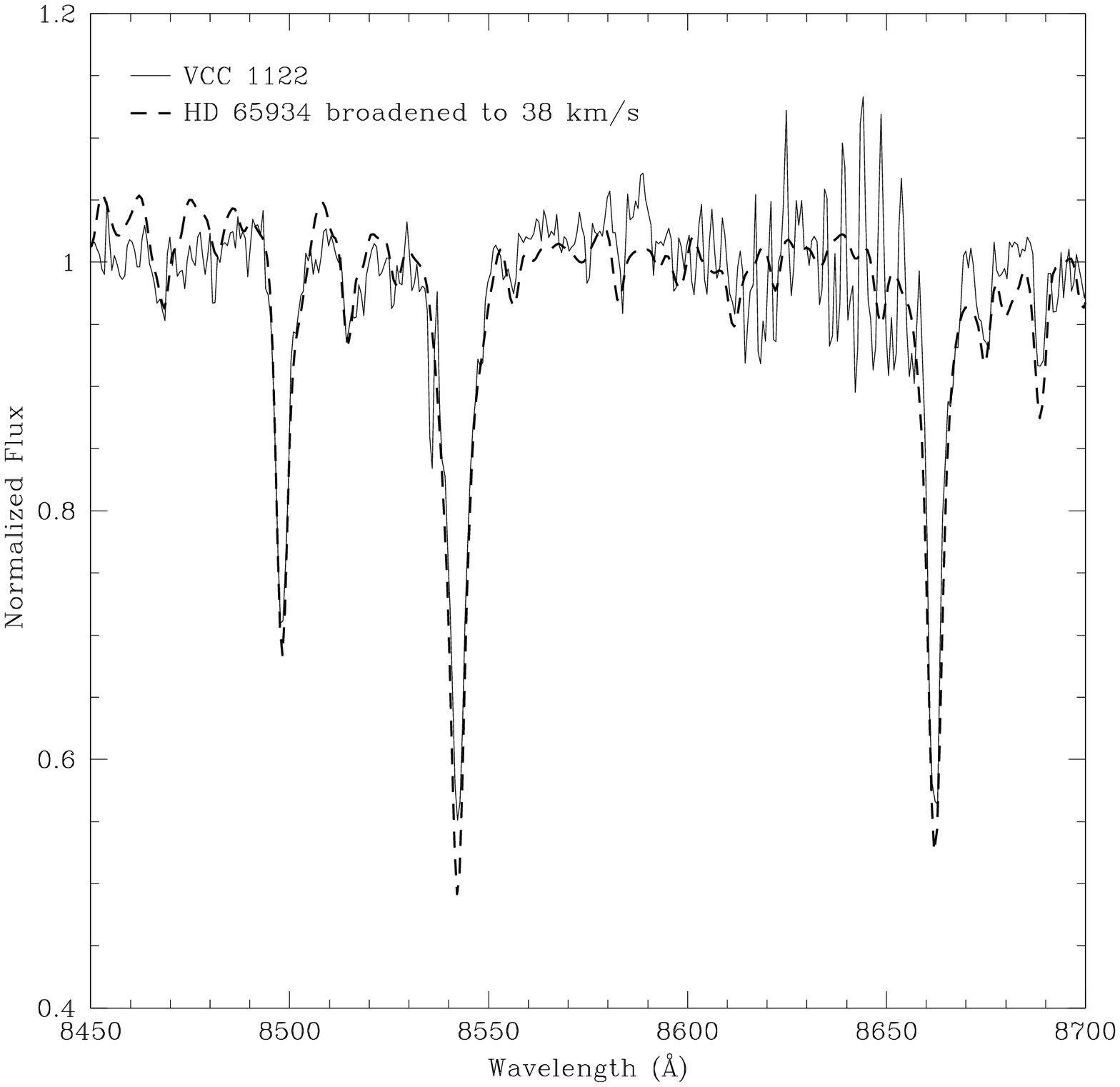,height=15.cm}
\figcaption[vanzee.fig4.ps]{Comparison of the doppler corrected Ca triplet line profiles 
of the integrated spectrum of VCC 1122 and those of a G8 III template star (HD 65934) 
broadened with a velocity dispersion of 38 \kms.  The width of the Ca lines of the template
star are well matched to the galaxy spectrum, confirming that the results from
the cross correlation analysis are properly calibrated.
\label{fig:comp}}

\psfig{figure=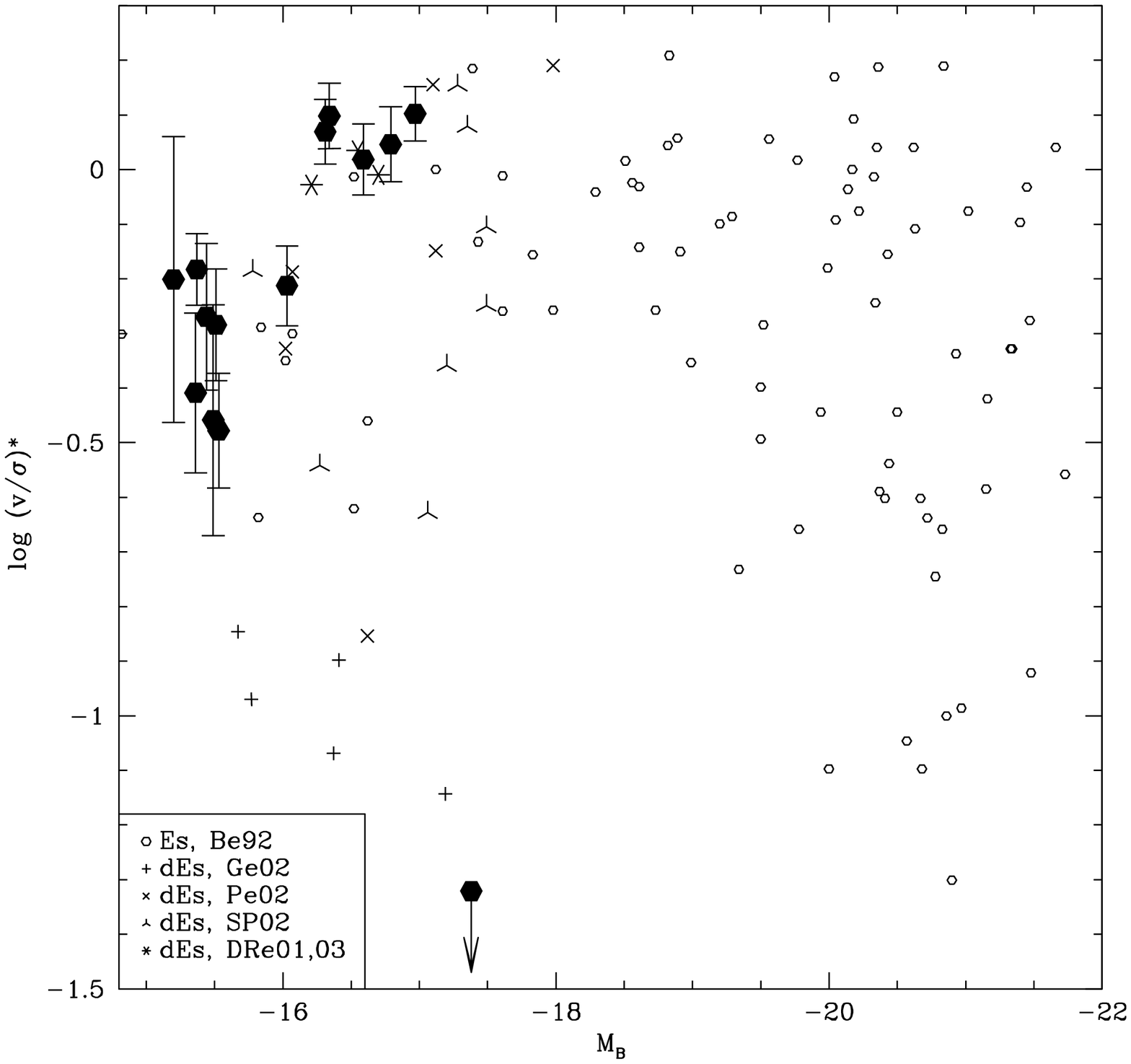,height=15.cm}
\figcaption[vanzee.fig5.ps]{The anisotropy parameter (v/$\sigma$)* vs absolute
magnitude for dEs and Elliptical galaxies.  Filled hexagons, plus signs,
crosses, triads, and
stars, represent the
dE galaxies from the present study, Geha et al. (2002), Pedraz et al. (2002),
Simien \& Prugniel (2002), and De Rijcke et al. (2001,2003) respectively.
The location of VCC 1261 is shown as an upper limit since the
observed velocity gradient was consistent with zero rotation.
Elliptical galaxies and dEs from Bender et al. (1992) are shown
with open circles.  Many of the dEs in the present sample are rotationally
flattened.  \label{fig:anisotrop} }

\psfig{figure=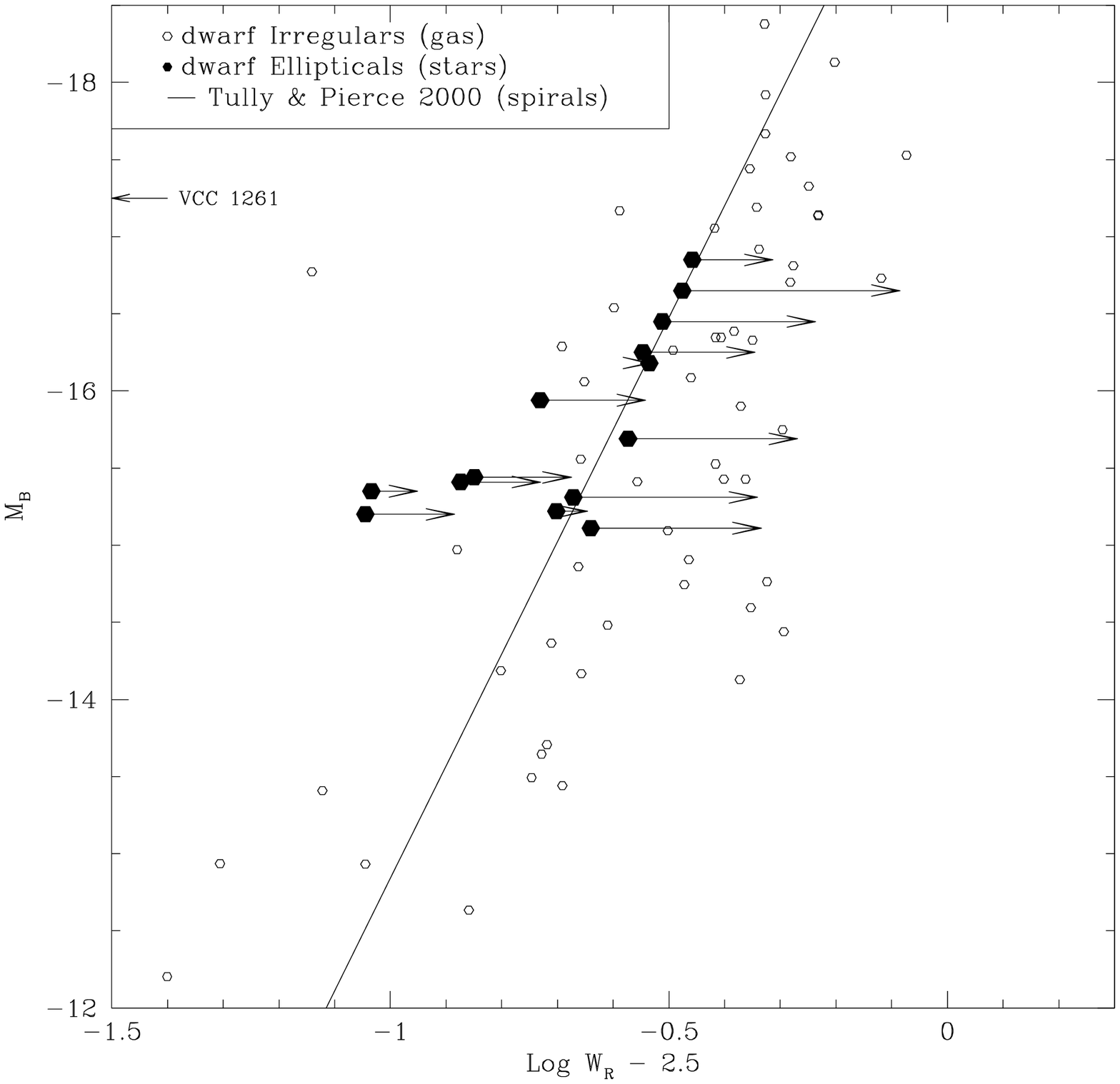,height=15.cm}
\figcaption[vanzee.fig6.ps]{Tully Fisher relation for dIs, dEs, and spirals. The
dE galaxies are shown as both filled hexagons (observed maximum rotation width)
and as arrows (maximum rotation width assuming the rotation curve could be
traced to 2.45 scale lengths).  The open circles are the dwarf irregular galaxies
compiled in van Zee (2001).  The straight line shows the expected relationship
for gas-rich spiral galaxies (Tully \& Pierce 2000). \label{fig:tf}}

\psfig{figure=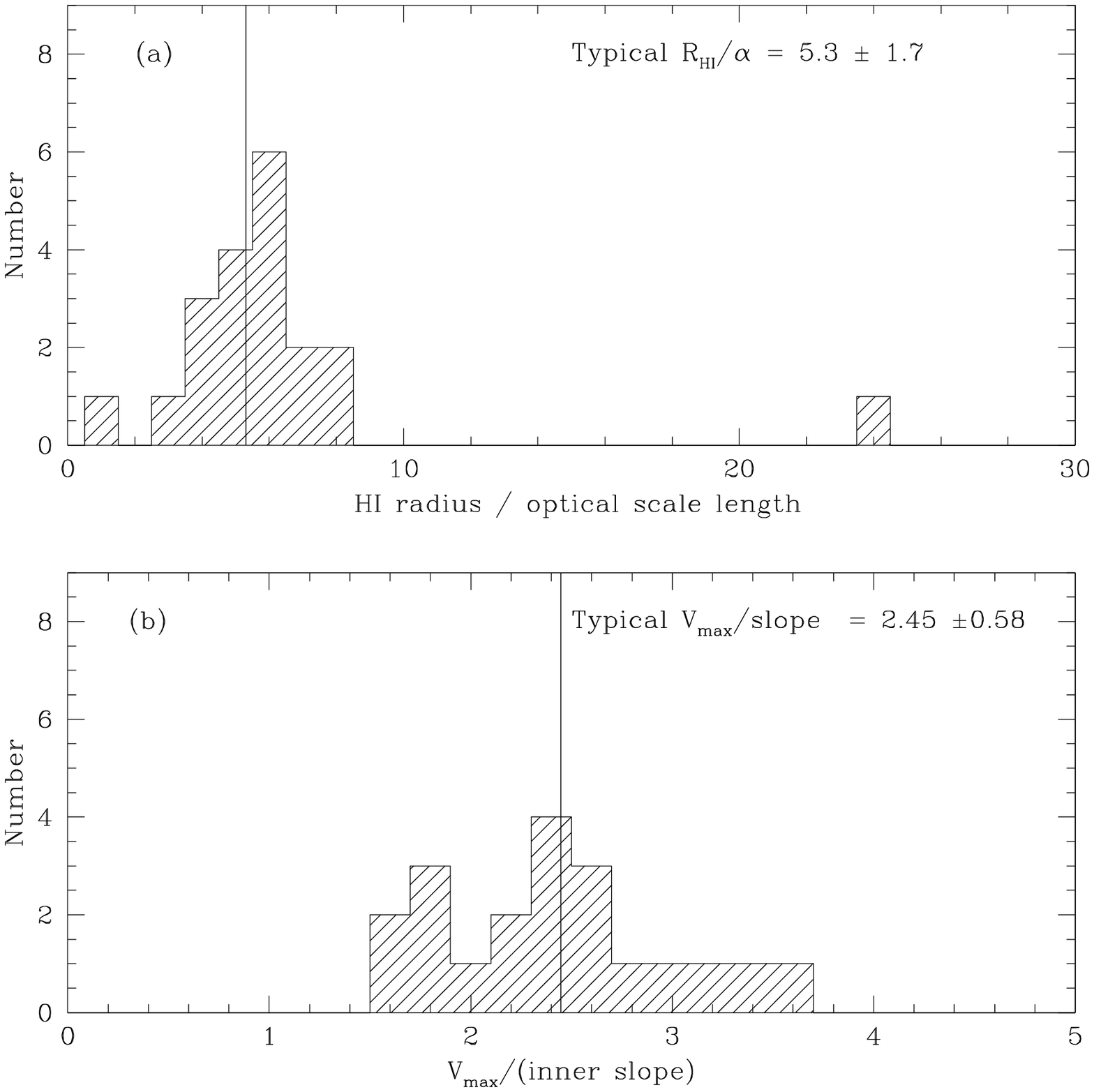,height=15.cm}
\figcaption[vanzee.fig7.ps]{(a) Histogram of the HI-to-optical size for dwarf
irregular galaxies with resolved neutral gas distributions and kinematics.
The typical rotation curve for a gas-rich dwarf galaxy can be traced out to 
5.3 times the optical scale length using neutral hydrogen.  (b) The ratio
of the maximum rotation velocity, V$_{\rm max}$, to the slope of the inner
rotation curve for gas-rich dwarf irregular galaxies.  The typical dwarf
irregular galaxy has a V$_{\rm max}$/slope of 2.45 scale lengths.
\label{fig:slope}}

\psfig{figure=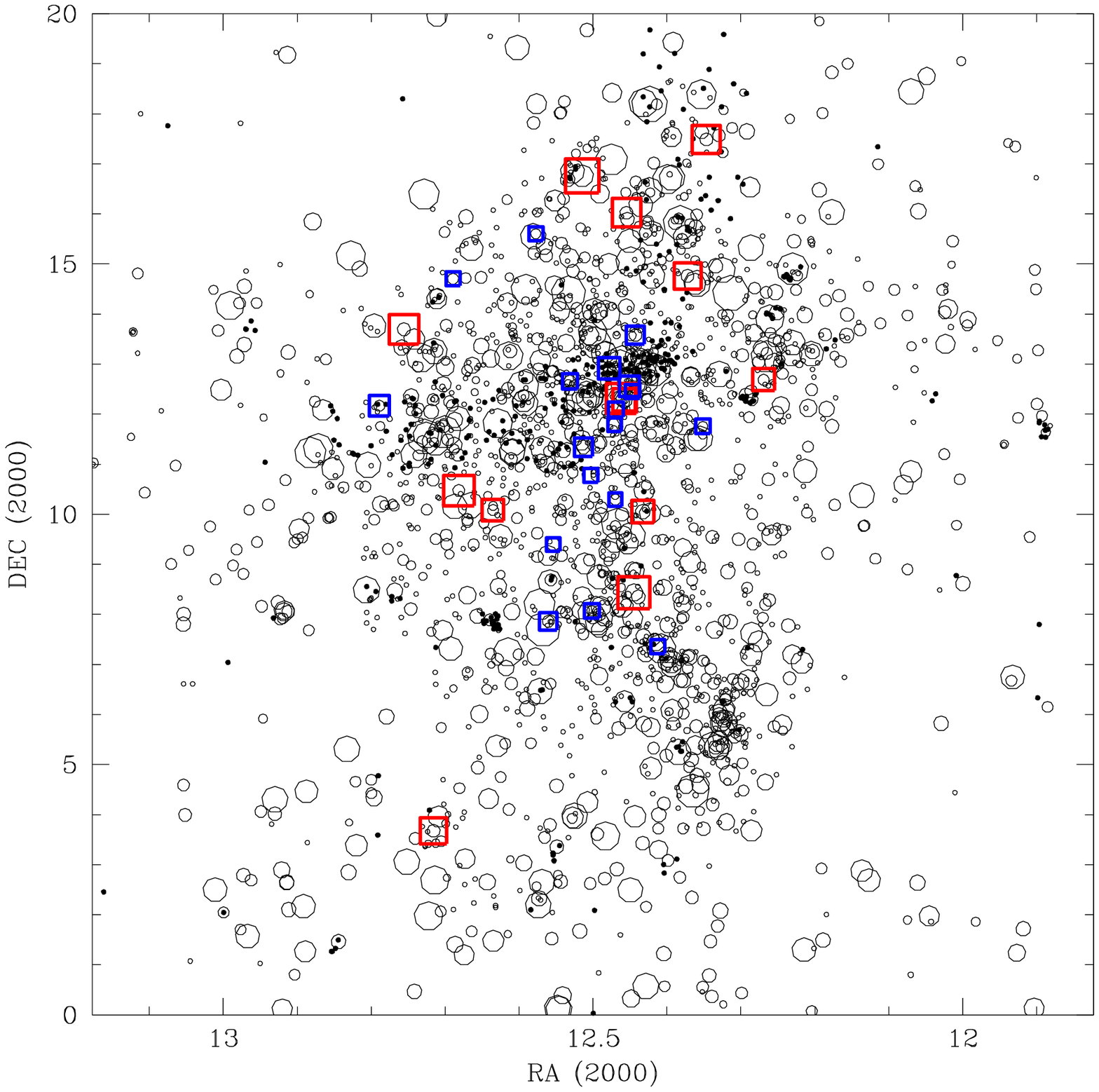,height=15.cm}
\figcaption[vanzee.fig8.ps]{
Location of the rotating (red squares) and non-rotating (blue squares)
dwarf elliptical galaxies in the Virgo cluster.  The relative size of the
square is indicative of the galaxy's anisotropy parameter (v/$\sigma$)*.
The complete kinematic sample includes those in Geha et al.\ 2003; Pedraz et al.\ 2002;
and this paper.  Also shown are all cataloged galaxies
believed to be in Virgo (circles).  Here, the symbol size is proportional
to the galaxy luminosity (open circles);  galaxies in the Virgo Cluster Catalog
with no listed apparent magnitudes are also included (small filled circles).
There is a slight trend for the non-rotating dwarf elliptical
galaxies to be located toward the cluster center or in the dense clumpy
regions.
\label{fig:cluster}}


\begin{thebibliography}{}
\bibitem[Bender et al.(1992)]{BBF92} Bender, R., Burstein, D., \& Faber, S. M. 1992, \apj, 399, 462
\bibitem[Bender \& Nieto(1990)]{BN90} Bender, R., \& Nieto, J.--L. 1990, \aap, 239, 97
\bibitem[Bender et al.(1991)]{BPN91} Bender, R., Paquet, A., \& Nieto, J.--L. 1991, \aap, 246, 349
\bibitem[Binggeli et al.(1993)]{BPT93} Binggeli, B., Popescu, C. C., \& Tammann, G. A., 1993, \aaps, 98, 275
\bibitem[Binggeli et al.(1985)]{VCC} Binggeli, B., Sandage, A. \& Tammann, G. A. 1985, \aj, 90, 1681
\bibitem[Binggeli et al.(1988)]{BST88} Binggeli, B., Sandage, A. \& Tammann, G. A. 1988, \araa, 26, 509
\bibitem[Binggeli et al.(1987)]{BST87} Binggeli, B., Tammann, G. A. \& Sandage, A. \& 1987, \aj, 94, 251
\bibitem[Binggeli, Tarenghi, \& Sandage(1990)]{BTS90} Binggeli, B., Tarenghi, M., \& Sandage, A.\ 1990, \aap, 228, 42
\bibitem[Conselice et al.(2003)]{COGW03} Conselice, C. J., O'Neil, K., Gallagher, J. S., \& Wyse, R. F. G. 2003, \apj, 591, 167
\bibitem[Conselice et al.(2001)]{CGW01} Conselice, C. J., Gallagher, J. S., \& Wyse, R. F. G. 2001, \apj, 559, 791
\bibitem[C\^ot\'e, Carignan, \& Freeman(2000)]{CCF00} C\^ot\'e, S., Carignan, C., \& Freeman, K. C. 2000, \aj, 120, 3027
\bibitem[Courteau \& Rix(1999)]{CR99} Courteau, S. \& Rix, H.-W. 1999, \apj, 513, 561
\bibitem[Davies \& Phillipps(1988)]{DP88} Davies, J. I., \& Phillipps, S. 1988, \mnras, 233, 553
\bibitem[De Rijcke et al.(2001)]{DDZH01} De Rijcke, S., Dejonghe, H., Zeilinger, W.~W., \& Hau, G.~K.~T.\ 2001, \apj, 559, L21
\bibitem[De Rijcke et al.(2003)]{DDZH03} De Rijcke, S., Dejonghe, H., Zeilinger, W.~W., \& Hau, G.~K.~T.\ 2003, \aap, 400, 119
\bibitem[De Young \& Heckman(1994)]{DH94} De Young, D. S., \& Heckman, T. M. 1994, \apj, 431, 598
\bibitem[De Young \& Gallagher(1990)]{DG90} De Young, D. S., \& Gallagher, J. S. 1990, \apj, 356
\bibitem[Dekel \& Silk(1986)]{DS86} Dekel, A., \& Silk, J. 1986, \apj, 303, 39
\bibitem[Drinkwater \& Hardy(1991)]{DH91}   Drinkwater, M., \& Hardy, E. 1991, \aj, 101, 94
\bibitem[Ferrara \& Tolstoy(2000)]{FT00} Ferrara, A., \& Tolstoy, E. 2000, \mnras, 313, 291
\bibitem[Ferguson \& Binggeli(1994)]{FB94}   Ferguson, H. C., \& Binggeli, B. 1994, \aapr, 6, 67
\bibitem[Ferguson \& Sandage(1989)]{FS89} Ferguson, H. C., \& Sandage, A. 1989, \apj, 346, 53
\bibitem[Gallagher \& Wyse(1994)]{GW94} Gallagher, J.~S.~\& Wyse, R.~F.~G.\ 1994, \pasp, 106, 1225
\bibitem[Geha et al.(2002)]{GGvM02} Geha, M., Guhathakurta, P., \& van der Marel, R. P. 2002, \aj, 124, 3073
\bibitem[Geha et al.(2003)]{GGvM03} Geha, M., Guhathakurta, P., \& van der Marel, R. P. 2003, \aj, 126, 1794
\bibitem[Giovanelli \& Haynes(1983)]{GH83} Giovanelli, R., \& Haynes, M. P. 1983, \aj, 88, 881
\bibitem[Gunn \& Gott(1972)]{GG72} Gunn, J.~E.~\& Gott, J.~R.~I.\ 1972, \apj, 176, 1
\bibitem[James(1994)]{J94}    James, P. A. 1994, \mnras, 269, 176
\bibitem[Jerjen et al.(2004)]{JBB04} Jerjen, H., Binggeli, B. \& Barazza, F.D. 2004, \aj, 127, 771
\bibitem[Kelson et al.(2000)]{Ke00} Kelson, D. D. et al.\ 2000, \apj, 529, 768
\bibitem[Kobulnicky \& Skillman(1995)]{KS95} Kobulnicky, H. A., \& Skillman, E. D. 1995, \apj, 454, L121
\bibitem[Koopmann \& Kenney(1998)]{KK98} Koopmann, R. A., \& Kenney, J. D. P. 1998, \apj, 497, L75
\bibitem[Kormendy(1985)]{K85}     Kormendy, J. 1985, \apj, 295, 73
\bibitem[Lin \& Faber(1983)]{LF83} Lin, D. N. C. \& Faber, S. M. 1983, \apjl, 266, L21
\bibitem[Loose \& Thuan(1986)]{LT86}   Loose, H.--H., \& Thuan, T. X. 1986, \apj, 309, 59
\bibitem[Mac Low \& Ferrara(1999)]{MF99}   Mac Low, M.--M., \& Ferrara, A. 1999, \apj, 513, 142
\bibitem[Marcolini, Brighenti, \& D'Ercole(2003)]{MBD03} Marcolini, A., Brighenti, F., \& D'Ercole, A. 2003, \mnras, 345, 1329
\bibitem[Mayer et al.(2001a)]{Me01a} Mayer, L., Governato, F., Colpi, M., Moore, B., Quinn, T., Wadsley, J., Stadel, J., \& Lake, G.\ 2001a, \apjl, 547, L123
\bibitem[Mayer et al.(2001b)]{Me01b} Mayer, L., Governato, F., Colpi, M., Moore, B., Quinn, T., Wadsley, J., Stadel, J., \& Lake, G.\ 2001b, \apj, 559, 754
\bibitem[Neilsen \& Tsvetanov (2000)]{NT00}  Neilsen, E.H. Jr. \& Tsvetanov, Z.I. 2000, \apj, 536, 155
\bibitem[Oh \& Lin (2000)]{Oh00}  Oh, K.S. \& Lin, D.N.C 2000, \apj, 543, 620
\bibitem[Oke(1990)]{oke90}  Oke, J. B. 1990, \aj, 99, 1621
\bibitem[Papaderos et al.(1996a)]{PLFT96} Papaderos, P., Loose, H.--H., Fricke, K. J., \& Thuan, T. X. 1996a, 
        \aap, 314, 59
\bibitem[Patterson \& Thuan(1992)]{PT92} Patterson, R.~J.~\& Thuan, T.~X.\ 1992, \apjl, 400, L55
\bibitem[Pedraz et al.(2002)]{PGCSBG02} Pedraz, S., Gorgas, J., Cardiel, N., S\'anchez-Bl\'azquez, P., \& Guzm\'an, R. 2002, \mnras, 332, L59
\bibitem[Richer \& McCall(1995)]{RM95} Richer, M. G., \& McCall, M. L. 1995, \apj, 445, 642
\bibitem[Ryden et al.(1999)]{RTPL99} Ryden, B. S., Terndrup, D. M., Pogge, R. W., \& Lauer, T. R. 1999, \apj, 517, 650
\bibitem[Sancisi, Thonnard, \& Ekers(1987)]{STE87} Sancisi, R., Thonnard, N., \& Ekers, R.~D.\ 1987, \apjl, 315, L39
\bibitem[Silich \& Tenorio-Tagle(1998)]{ST98}   Silich, S. A., \& Tenorio--Tagle, G. 1998, \mnras, 299, 249
\bibitem[Silk et al.(1987)]{SWS87}  Silk, J., Wyse, R. F. G., Shields, G. A. 1987, \apjl, 322, L59
\bibitem[Simien \& Prugniel(2002)]{SP02} Simien, F.~\& Prugniel, P.\ 2002, \aap, 384, 371
\bibitem[Skillman \& Bender(1995)]{SB95} Skillman, E. D., \& Bender, R. 1995, Rev.\ Mex.\ AASC, 3, 25
\bibitem[Skillman, Kennicutt, \& Hodge(1989)]{SKH} Skillman, E. D., Kennicutt, R. C., \& Hodge, P. W. 1989, \apj, 347, 875
\bibitem[Smecker-Hane et al.(1994)]{SHSHL94} Smecker-Hane, T. A., Stetson, P. B., Hesser, J. E., \& Lehnert, M. D. 1994, \aj, 108, 507
\bibitem[Solanes et al.(2001)]{SMGGGH01} Solanes, J.-M., Manrique, A., Garc\'ia-G\'omez, C., Gonz\'alez-Casado, G., Giovanelli, R., \& Haynes, M. P. 2001, ApJ, 548, 97
\bibitem[Solanes et al.(2002)]{SSSGH02} Solanes, J.-M., Sanchis, T., Salvador-Sol\'e,
        Giovanelli, R., \& Haynes, M. P. 2002, \aj, 124, 2440
\bibitem[Sung et al.(1998)]{SHRCK98} Sung, E.--C., Han, C., Ryden, B. S., Chun, M.--S., \& Kim, H.--I.
        1998, \apj, 499, 140
\bibitem[Tonry \& Davis(1979)]{TD79} Tonry, J., \& Davis, M. 1979, \aj, 84, 1511
\bibitem[Tully \& Pierce(2000)]{TP00} Tully, R. B., \& Pierce, M. J. 2000, \apj, 533, 744
\bibitem[Tully \& Shaya(1984)]{TS84} Tully, R.~B.~\& Shaya, E.~J.\ 1984, \apj, 281, 31
\bibitem[van den Bergh(1994)]{vdB94} van den Bergh, S.\ 1994, \apj, 428, 617
\bibitem[van Zee(2001)]{vZ01}  van Zee, L. 2001 \aj, 121, 2003
\bibitem[van Zee \& Barton(in prep)]{vZB04} van Zee, L., \& Barton, E. G. 2004, in prep
\bibitem[van Zee et al.(1997a)]{vHS97} van Zee, L., Haynes, M. P., \& Salzer, J. J. 1997a, \aj, 114, 2479
\bibitem[van Zee et al.(1997b)]{vHSB97} van Zee, L., Haynes, M. P., Salzer, J. J., \& Broeils, A. H. 1997b,
       \aj, 113, 1618
\bibitem[van Zee et al.(2001)]{vSS01}  van Zee, L., Salzer, J. J., \& Skillman, E. D. 2001, \aj, 122, 121
\bibitem[West \& Blakeslee(2000)]{WB00}  West, M.J. \& Blakeslee, J.P. 2000, \apj, 543, L27
\bibitem[Yasuda et al. (1997)]{YFO97} Yasuda, N., Fukugita, M. \& Okamura, S. 1997, \apjs, 
108, 417
\bibitem[Young \& Currie(1995)]{YC95}   Young, C. K. \& Currie, M. J. 1995, \mnras, 273, 1141
\end{thebibliography}
\end{document}